\documentclass[]{aa}
\usepackage{graphicx}
\usepackage{rotating,subfigure}
\usepackage{subfigure}
\def\xmm{{\it XMM-Newton\/}}
\def\etal{et al.\ }
\def\betamod{$\beta$-model}
\def\mgas{$M_{\rm gas}$}
\def\fgas{$f_{\rm gas}$}
\def\kT {{\rm k}T}
\def\Sx {S_{\rm X}}
\def\Lx {L_{\rm X}}
\def\Tx {T}
\def\Mv {M_{\rm V}}
\def\keV {\rm keV}
\def\rs {r_{\rm s}}
\def \rv {r_{200}}
\def\rc {r_{\rm c}}
\def \ne {n_{\rm e}}

\def \lsun {\rm h_{50}^{-2}~L_{\odot}}

\def \lb {L_{\rm B}}
\def \lbj {L_{\rm B_{\rm j}}}

\def \mlb {M/L_{\rm B}}
\def \mlv {M/L_{\rm V}}

\def \mlsun {\rm h_{50}~M_{\odot}/L_{\odot}}
\def \mlfsun {\rm h_{50}^{-1/2}~M_{\odot}/L_{\odot}}

\begin{document}
     \title{Entropy scaling in galaxy clusters: insights from an \xmm\
     observation of the poor cluster \object{A1983}}

     \author{G. W. Pratt  \and M. Arnaud
     }
     \offprints{G. W. Pratt, \email{gwp@discovery.saclay.cea.fr}}

     \institute{CEA/Saclay, Service d'Astrophysique,
                L'Orme des Merisiers, B\^{a}t. 709,
                91191 Gif-sur-Yvette Cedex, France
               }
     \date{ }

\abstract{ An \xmm\ observation of the cool ($\kT=2.1~\keV$) cluster
A1983, at $z=0.044$, is presented. Gas density and temperature
profiles are calculated over the radial range up to $500~{\rm
h_{50}^{-1}~kpc}$, corresponding to $\sim 0.35~\rv$.  The outer
regions of the surface brightness profile are well described with a
\betamod\ with $\beta=0.74$, but the central regions require the
introduction of a second component.  The temperature profile is flat
at the exterior with a slight dip towards the centre.  The total mass
profile, calculated from the temperature and density information
assuming hydrostatic equilibrium, is consistent with an NFW profile,
but with a low concentration parameter $c=3.75 \pm 0.74$, which
may be due to the cluster not being totally relaxed.  
Published optical data are used to calculate the $\mlb$ ratio profile and
the overall iron mass over luminosity ratio.
The $\mlb$ ratio profile shows that, at large scale, light traces
mass to a reasonable extent, and the $\mlb$ ratio at $0.35 r_{200}$ 
($\mlb = 135 \pm 45\ h_{50} M_{\odot}/L{\odot}$) is consistent with 
the trends with mass observed in the optical. 
The iron mass over luminosity ratio
is about two times less than that observed for a cluster at 5 keV. The gas mass
fraction rises rapidly in the central regions to level off quickly at
$\sim 200~{\rm h_{50}^{-1}~kpc}$; the value at $0.35~\rv$ is $\sim
8\%$.  The scaling properties of the emission measure profile are
consistent with the empirical relation \mgas$ \propto \Tx^{1.94}$; use
of the standard self-similar relation \mgas$ \propto \Tx^{1.5}$
results in a scaled profile that is a factor of about two too low as compared
to the reference mean profile for hot clusters.
Comparison of the entropy profile of this cool cluster with that of
the hot cluster A1413 shows that the two profiles are extremely well
scaled using the empirically determined relation $S \propto
\Tx^{0.65}$, suggesting that the slope of the $S$--$\Tx$ relation is
shallower than expected in the standard self-similar model.  The form
of the two entropy profiles is remarkably similar, and there is no
sign of a larger isentropic core in the cooler cluster. 
These data provide powerful agruments against preheating models.
In turn, there is now  increasing observational support for a trend of 
\fgas\ with system mass, which may go some way towards explaining the
observed scaling behaviour. 

 }

     \authorrunning{G.W. Pratt \& M. Arnaud}
     \titlerunning{An \xmm\ observation of \object{A1983}} \maketitle
%

\section{Introduction}
In the simplest models of gravitational collapse (e.g.,
Bertschinger~\cite{bert85}), supported by numerical simulations without
non-gravitational gas processes (e.g., Navarro, Frenk \&
White~\cite{nfw97} [NFW]) and analytical models (e.g., Cavaliere,
Menci \& Tozzi~\cite{cmt99}), the intrinsic properties of galaxy clusters
(e.g., X-ray luminosity and temperature $\Lx, \Tx$, virial mass $\Mv$)
follow self-similar scaling derived, assuming HE, from basic virial
relations.  Such self-similarity applies equally to the hot
intracluster medium (ICM) and the dark matter components.  In
particular, $M_{\rm \delta} \propto \Tx^{3/2}$,
where $M_{\rm \delta}$ is the total mass in a sphere of
radius $R_{\rm \delta}$ corresponding to the overall density contrast
$\delta$, and $\Tx$ is the cluster X-ray temperature.  The virialised
part of a cluster roughly corresponds to $\delta=200$ (e.g., Evrard,
Metzler \& Navarro~\cite{emn96}).  Assuming a constant gas mass
fraction \fgas = \mgas$/\Mv$, the total gas mass then scales as
\mgas$\propto \Tx^{3/2}$.  Furthermore, if the X-ray emission is
dominated by Bremsstrahlung, then $\Lx \propto \Tx^2$.

However, it has been known for more than a decade that real clusters
do not follow these laws (e.g., Edge \& Stewart~\cite{es91}).  The
most notable example is perhaps the $\Lx$--$\Tx$ relation, with
observations suggesting $\Lx \propto \Tx^{\sim 2.9}$ (e.g., Arnaud \&
Evrard~\cite{ae99}).  In the study of why observed cluster properties
deviate from the expected scalings, it is the $\Lx$--$\Tx$ relation
which has been in the spotlight, mainly because this relation deals
with easily observable global properties.  Other important scaling
relations exist, however. Of particular interest for the present work
are the \mgas--$\Tx$ and entropy $S$--$\Tx$ relations, because both of
these relations probe variations in gas content and 
can directly be derived from the data.

Studies of large samples, such as those by Neumann \& Arnaud
(\cite{na01}) and Mohr, Mathiesen \& Evrard (\cite{mme99}) have shown
that, in fact, \mgas$ \propto \Tx^{1.9}$, a steepening which can be
interpreted as a dependence of \fgas\ {\bf on} temperature.  However, the
direct calculation of \fgas\ is fraught with difficulties, requiring a
robust method for calculating the total mass $\Mv$, and so whether \fgas\
does depend on temperature is still under debate (e.g., Arnaud \&
Evrard~\cite{ae99}, Mohr \etal~\cite{mme99}, Roussel, Sadat \&
Blanchard~\cite{rsb00}).

The entropy is a
formidable tool for studying the thermodynamic history of the ICM 
(e.g., Voit et al~\cite{voitetal03}).
Defining the `entropy' as $S = T/\ne^{2/3}$ (e.g., Lloyd-Davies,
Ponman \& Cannon~\cite{ldpc00}), self-similarity implies a simple
scaling of the entropy with temperature, such that $S \propto \Tx$.
Ponman, Cannon \& Navarro (\cite{pcn99}) and Lloyd-Davies \etal
(\cite{ldpc00}) used ROSAT observations directly to detect entropies
exceeding levels attainable through gravitational collapse alone, in
the inner regions (0.1 $\rv$) of low mass systems.  From their sample
of 20 systems, it appeared that the entropy followed the expected
self-similar scaling down to a certain floor level of $140~ {\rm
h_{50}^{-1/3}}~\keV~{\rm cm^{2}}$, implying that non-gravitational
processes act to set a lower limit to the entropy
possible for the gas in collapsed haloes.  However, very recent results
with a much larger sample of 66 systems (Ponman, Sanderson \&
Finoguenov~\cite{psf03}), indicate that, in fact, there exists a slope
in $S(T)$ which is significantly shallower than the self-similar
relation throughout, such that $S \propto T^{\sim 0.65}$.  This again
calls into question the fundamental assumption of a constant \fgas.

The commonest explanation for the deviation from simple self-similar scaling
requires non-gravitational energy input, raising the gas to a higher
entropy level at some time in its history.  However, there is little
consensus regarding the source of this extra entropy.  Pre-heating has
been suggested (originally by Kaiser~\cite{kaiser91} and Evrard \&
Henry~\cite{evrard91}), although there is no agreement on the
astrophysical source responsible for it: early galactic winds (e.g.,
Loewenstein~\cite{loewenstein}), AGN (Valageas \&
Silk~\cite{valageas}), or possibly both.  Other effects may also play
a role, like radiative cooling (e.g., Muanwong \etal~\cite{muan01}),
or variation of galaxy formation efficiency with system mass
(Bryan~\cite{bryan00}).

Cool clusters are ideal targets for investigating the entropy 
because in such systems, the excess entropy becomes more
evident as the processes which produce it affect the ICM in direct
competition with standard gravitational heating.  A1983 is well
studied in optical, with $\sim 125$ galaxy velocities measured in the
field by Dressler \& Shectman (\cite{ds88}).  Girardi \etal
(\cite{getal93}) derive a velocity dispersion of $551^{+71}_{-47}$ km~s$^{-1}$
from these data, a relatively low value which implies that
A1983 is a cool/poor cluster.

In contrast, A1983 is little studied in X-ray.  It was observed with
Einstein (Jones \& Forman~\cite{jf84}), giving $\Lx\ [0.5-3.0]\ \keV =
2.12\pm0.44 \times 10^{43}$ erg~s$^{-1}$, and has been detected in the ROSAT
All Sky Survey (Ebeling \etal~\cite{ebeling}) with a quoted $\Lx\
[0.1-2.4\ \keV] = 0.47 \times 10^{44}$ erg~s$^{-1}$.  These low $\Lx$ values
again point to the poor nature of the cluster.

This paper first presents results from the new \xmm\ observation of
A1983. The entropy and emission measure profiles of this cool cluster
are then compared with those of hot clusters.
These are two emphases: a direct comparison in form between
the radial profiles, which shows the unparalleled quality of these
\xmm\ data; and how the scaling relations involving these quantities
varies across the temperature range between the clusters. This paper
will show the remarkable similarity in form between the 
profiles which is difficult to reconcile with conventional preheating
models, and will use the \mgas--$\Tx$ and $S$--$\Tx$ relations to
argue that the gas mass fraction \fgas\ cannot be constant with
cluster mass.

Taking $H_0 = 50$ km s$^{-1}$ Mpc$^{-1}$ and $\Omega_0 =1$, 
1\arcmin\ corresponds to 71.4 kpc at the cluster redshift of 0.044.


\section{Data preparation and analysis}

\subsection{Preliminaries}

The \xmm\ observations of A1983 were taken in Revolution 400, on 2002
February 14, and have a total exposure time of $\sim 32$ ks. Pipeline
products provided by the \xmm\ SOC, consisting of calibrated event
files preprocessed with the SAS, are used in this analysis. All
camera exposures used the THIN1 filter. The MOS exposures were taken
in standard Full Frame mode, and for these files PATTERNs 1-12 were
considered. The pn observations were obtained in Extended Full
Frame (EFF) mode. In view of the calibration uncertainties associated
with the EFF mode, only PATTERN 0 was considered, corresponding to the
well-calibrated single events.

The dedicated blank-sky event files accumulated by Lumb (\cite{lumb})
were used as background files.  Events were extracted from each event
file according to the same PATTERN selection as used for the source
observations.  The background files were then recast into the
coordinates of the source observation, ensuring that
source and background products are extracted from the same regions of
the detector, thus minmising detector variations.

To ensure that all output products (spectra, surface brightness
profiles, images) are corrected for vignetting, the photon weighting
method of Arnaud \etal (\cite{arnaudetal01a}) was used.  This is
applied to both source and background data sets.  Note that while the
background component induced by cosmic rays is not vignetted, source
and background event files are treated in the same way, so that the
correction factor is the same and thus cancels.


\subsection{Flare removal}

\begin{figure}
\begin{centering}
\includegraphics[scale=0.5,angle=0,width=\columnwidth,keepaspectratio]{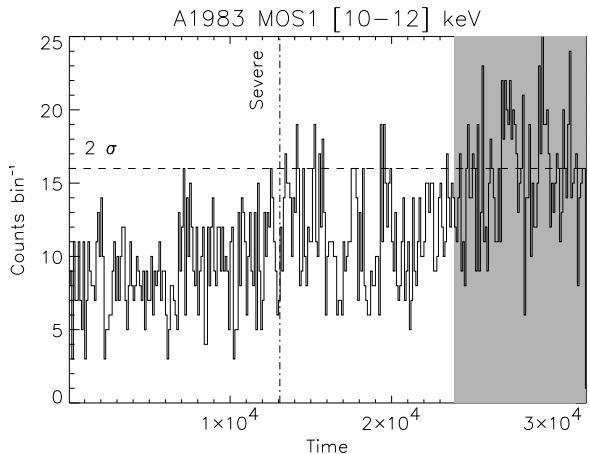}

\caption{{\footnotesize The MOS1 light curve in the [10-12] keV band.
All frames in the shaded area were excluded because of flare
activity. The 2$\sigma$ cutoff level (dotted line), established after
excluding the shaded region, and the frames used for the ``severe''
analysis (dot-dash line), are also shown. Further details in the
text.}}\label{fig:fig1}
\end{centering}

\end{figure}

Inspection of the MOS light curves in the source-free [10-12] keV
band (Fig.~\ref{fig:fig1}) shows that the background level during
this observation is not constant. The MOS [10-12] keV light curves
are very similar to the pn [12-14] keV light curve. These high energy
curves show that the overall level seems to be slowly increasing,
with some low-level flaring activity appearing near the end of the
observation. The fluctuations are not large in these bands, but could
potentially have an effect on spectra in the outer regions of the
cluster, where the emission begins to be dominated by the background.

However, inspecting the light curves at lower energies, it is obvious
that, for this observation at least, any criterion applied in the
high energy band would miss some flares which appear to reach peak
amplitude in lower energy bands.

It is prudent to exclude all frames near the end of the observation
(the last $\sim 5$ ks), where flaring begins to become important. To test
the effect of imposing different flare cutoff thresholds, three event
files were then produced, filtered at different levels of severity,
increasingly diminishing the final exposure times. The background
event lists were treated using exactly the same criteria.

\begin{enumerate}
\item In the ``$2 \sigma$'' event list, using the [10-12] keV light
curve, frames were excluded which were outside a $2 \sigma$ threshold
according to the method outlined in Pratt \& Arnaud  (\cite{pa02}).  The
final exposure times are 23/16 ks (MOS/pn).  \item There are some
observations where low-energy flares exist which do not appear at high
energies.  Light curves were thus extracted in several bands.  For the
MOS detectors, these bands were [0.3-1.4], [2.0-5.0], [5.0-8.0],
[8.0-10.0] and [10.0-12.0] keV. For the pn, the [8.0-10.0] keV band
was ignored because of strong instrumental lines, but a further light
curve was extracted from events with energies between [12.0-14.0] keV.
The $2 \sigma$ filtering described above was then applied to each
light curve, generating 5 GTI files.  The observation was then
filtered using the intersection of all these GTI files for a final
``intersect'' event list.  The resulting exposure times are 18/12 ks
(MOS/pn).  \item The first $\sim 13$ ks of the observation appears to
be relatively quiescent, and has a mean count rate similar to that of
the background files.  A ``severe'' event list was generated by
discarding all frames after this period, leaving 13/8 ks (MOS/pn) of
useful data.
\end{enumerate}

After masking extraneous sources, surface brightness profiles were
extracted from each event list in several different bands. These
profiles were background subtracted and compared. They are identical
within their errors apart from differences in normalisation. To
assess the effect of the filtering on spectral products, spectra were
extracted in annular regions, centred on the peak of the X-ray
emission from the cluster (this is is discussed in more detail
below). The annular spectra from each event file were fitted with a
MEKAL model at the redshift of the cluster, absorbed with the
galactic absorption. MOS1, MOS2 and pn files were fitted both
independently and simultaneously. The profiles from each event file
were again consistent within their errors.

It thus appears that the different flare filtering levels have little
or no effect on the main output products, likely due to the relatively
weak nature of the fluctuations.  It is nevertheless important to
establish this, as incomplete flare screening has been suggested as a
source of discrepancy between \xmm\ and {\it Chandra\/} observations of
A1835 (Markevitch~\cite{mark02}).  In an effort to strike a delicate
balance between enthusiasm and caution, the ``intersect'' event lists
are thus adopted for the remainder of this analysis.


\subsection{Background subtraction}
\label{sec:bgsub}
For each of the output products (spectra and surface brightness
profiles), an equivalent product is extracted from the corresponding
blank-sky event list.  As discussed extensively in Pratt \etal
(\cite{pa01}) and Arnaud \etal (\cite{aml02}), while the blank-sky
observations provide an adequate description of the hard X-ray and
instrumental components of the \xmm\ background, they do not
necessarily represent the soft X-ray component of the background,
because this is variable across the sky (see Snowden
\etal~\cite{snow97}).  The method described in Pratt \etal
(\cite{pa01}) and Arnaud \etal (\cite{aml02}) is used to correct for
the difference of the background at low energy.  This correction is
applied to all spectra and surface brightness profiles discussed
below.

The cosmic ray background is variable at the $\sim 10\%$-level, and
so it is frequently necessary to normalise the background to the
level of the observation. The count rate in the [10-12] keV and
[12-14] keV (pn) bands, for MOS and pn respectively, is used in this
analysis, with each camera treated separately.



\section{Gas density distribution}

\subsection{Morphology}

\begin{figure}
\begin{centering}
\includegraphics[scale=1.,angle=-90,width=\columnwidth,keepaspectratio]{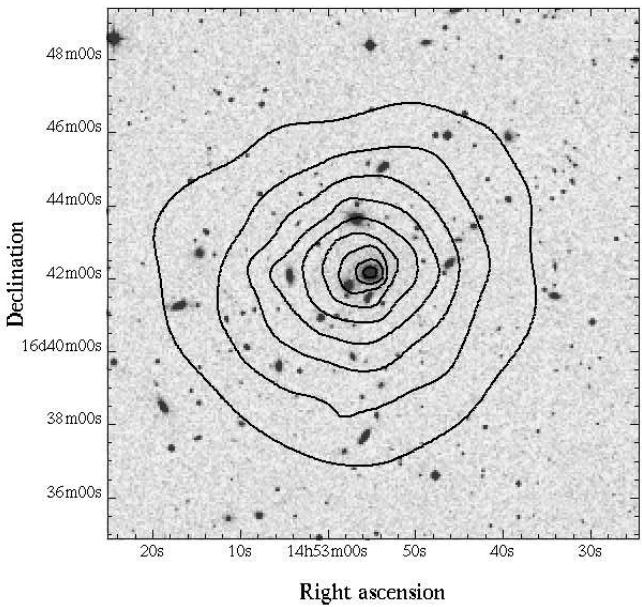}

\caption{{\footnotesize An XMM/DSS overlay image. The X-ray contours
are from an adaptively smoothed [0.3-2.0] keV non-background
subtracted MOS+pn image, and are logarithmically spaced by 0.15 dex
with the lowest contour at $4.2 \times 10^{-3}$ cts~s$^{-1}$~arcmin$^{2}$.
}}\label{fig:amas}
\end{centering}

\end{figure}

An XMM/DSS overlay image is shown in Fig.~\ref{fig:amas}.  The cluster
exhibits morphologically symmetric X-ray emission, centred
exactly on the early type galaxy \object{[WCB96] ACO 1983 B}, at
$z=0.04405$ (Wegner \etal~\cite{WCB96}; \cite{WCS99}).  Note that this
galaxy is not the Brightest Cluster Galaxy as defined by Postman \&
Lauer (\cite{PL}), which is located $12\arcmin$ further away in the
North, well beyond the X--ray emission, but the second ranked galaxy:
\object{[PL95] ACO 1983 G2}.

\subsection{Surface brightness profile}

Tha gas density profile can be derived from the surface brightness
profile, knowing the emissivity in the considered energy band.  This
is however more tricky for cool clusters than for hot clusters.  For
clusters like A1983 ($\kT\sim2~\keV$), the exponential cut-off of the
bremsstrahlung emission lies at low energies ($E\sim\kT$).
Furthermore, line emission is prominent, specially the FeL blend (see
Fig.~\ref{fig:globspec}) and radiative recombination radiation
contributes significantly to the continuum.  The emissivity,
$\Lambda$, in any considered energy band thus depends sensitively on
the abundance and temperature.  As will be seen below (see
Sect.~\ref{sec:temprof} and Sect.~\ref{sec:abprof}) radial variations of these
quantities exist in the cluster centre.  Thus, the
corresponding $\Lambda$ profile
has to be taken into account in the derivation of the gas density profile
from the surface brightness.  The fact that the temperature
and abundance profiles are determined with a cruder spatial resolution
than the $\Sx$ profile introduces a second complication and is a
potential source of systematic uncertainties. The surface brightness profile
was thus extracted in the [0.3-3.0] keV band
with the [0.9-1.2] keV band excluded to minimise the contribution from
the FeL blend, the feature which is the most sensitive to abundance
and temperature.

For each camera, the azimuthally averaged surface brightness profile
of the cluster and corresponding background was produced, after
masking of bright sources identified in the SAS pipeline detection.
The profiles were made directly from the event files by binning
weighted events into circular annuli about the emission peak.
The background subtraction, performed as described above in
Sect~\ref{sec:bgsub}, was undertaken separately for each camera.  The
resulting profiles are in excellent agreement with each other (bar the
expected normalisation factor), and so the MOS and pn profiles were
coadded.  The total profile was then rebinned so that a $S/N$ ratio of
$3\sigma$ was reached.

The background subtracted surface brightness profile
was then corrected for the remaining radial variation of $\Lambda$.
For that purpose a parametric model (in the form of
Eq.~\ref{eqn:allen}) was used for
the derived abundance and temperature profiles.
The corresponding emissivity profile $\Lambda(\theta)$ was then computed
in the considered energy band using an absorbed
thermal model convolved with the instrumental response.  The observed
surface brightness profile was then divided by $\Lambda(\theta)$
normalised to its value at large radii.  This correction is about
$50\%$ for the central bin, but drops to about $10\%$ at $\theta\sim
0.3'\sim 20$~kpc, and is negligable beyond $\sim 1\arcmin$.

\begin{figure}[t]
\begin{centering}
\includegraphics[scale=1.,angle=0,width=\columnwidth,keepaspectratio]{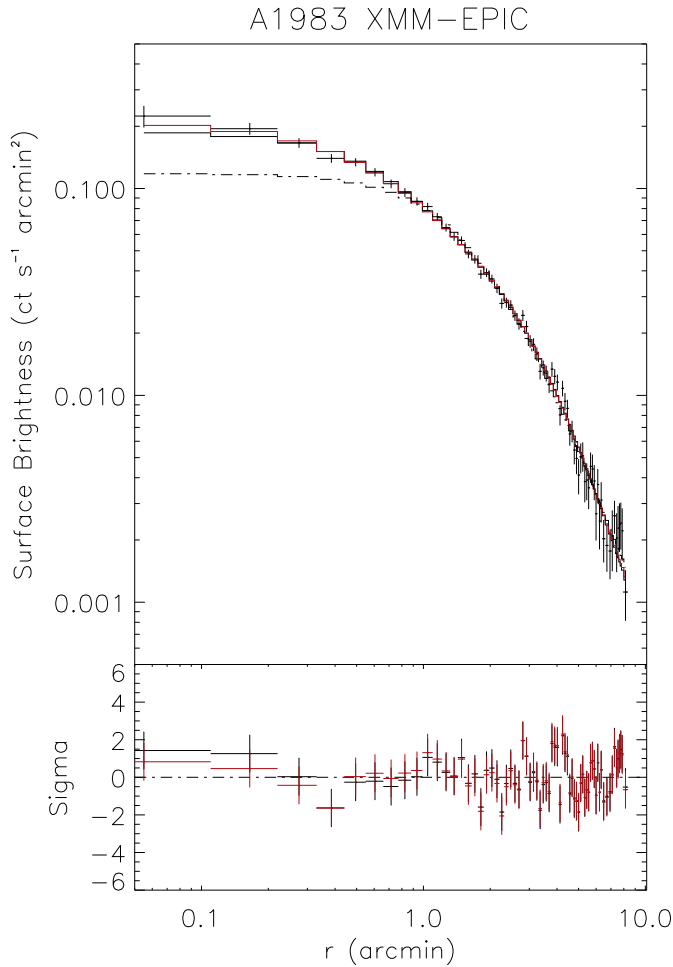}
\caption{{\footnotesize The combined EPIC surface brightness profile
of A1983 in the [0.3-3.0] keV band, with the [0.9-1.2]  keV band
excluded and corrected for the dependence of the emissivity,
$\Lambda$, on the temperature and abundance.  The profile is
background subtracted and corrected for vignetting.  Dot-dash line:
best fit \betamod\ to the outer ($\theta > 1\arcmin.0$) cluster
region.  The black and red lines are the best fit BB and KBB models
convolved with the \xmm\ PSF. See Sect.~\ref{sec:gasden} for model
details and Table~\ref{tab:sbfit} for the best-fit parameter
values.}}\label{fig:sbfits}
\end{centering}

\end{figure}

This corrected $\Sx$ profile is shown in Fig.~\ref{fig:sbfits}.
Cluster emission is significantly detected out to $8\arcmin.4$ or
$\sim 600$ kpc.  It is now directly proportional to the emission
measure profile, $EM(r)=\int_{r}^{\infty}\ne^{2}~dl$, where $n_{\rm
e}$ is the electronic density, which can be derived using the
emissivity at large radii.

\begin{table}
\begin{minipage}{\columnwidth}
\centering
\center
\caption{{\small Results of analytical fits to the gas surface
brightness profile, errors are $90\%$ confidence.}}
\begin{tabular}{ l l l}
\hline
\hline

\multicolumn{1}{l}{ Parameter }  & \multicolumn{1}{l}{ KBB model } &
\multicolumn{1}{l}{ BB model } \\
\multicolumn{1}{l}{ } & \multicolumn{1}{l}{ } & \multicolumn{1}{l}{ }
\\

\hline

$n_{\rm H,0} ({\rm h_{50}^{1/2}~cm^{-3}})$ & $6.64 \times 10^{-3}$ & $4.66
\times 10^{-3}$ \\
$r_{\rm c}$ & $2\arcmin.42^{+1.03}_{-0.28}$ &
$2\arcmin.65^{+0.73}_{-0.56}$ \\
$\beta$    & $0.73^{+0.16}_{-0.05}$ & $0.76^{+0.13}_{-0.08}$    \\
$R_{\rm cut}$   & $2\arcmin.16^{+1.19}_{-0.10}$&
$2\arcmin.40^{+0.55}_{-0.54}$ \\
$r_{\rm c, in}$ & $0\arcmin.40{\footnote{The maximum value of $r_{\rm
c, in}$ is fixed to $1\arcmin$.}}_{-0.15}$ &
$0\arcmin.49^{+0.16}_{-0.19}$ \\
$\xi$  & $0.63^{+0.29}_{-0.34}$  & -         \\
$\chi^2$/d.o.f       & 70.36/63 & 70.71/64  \\
$\chi^2_{\nu}$    & 1.12 & 1.10  \\

\hline
\end{tabular}
\label{tab:sbfit}
\smallskip

\end{minipage}
\end{table} %


\subsection{Gas density profile}
\label{sec:gasden}

It is convenient, especially for the total mass determination below
(Sect.~\ref{sec:mass}), to have an analytical description of the gas
density profile at all radii. The surface brightness profile
$S(\theta)$ was thus fitted with various parametric models, all of
which were convolved with the \xmm\ PSF and binned in the same way as
the observed profile.

A standard \betamod\ is a good description of the outer regions, 
but fails
nearer the centre ($\theta < 1\arcmin$), where the data show a slight
excess above that predicted from the fit.  Progressively cutting the
central region decreases the reduced $\chi^2$.  For the single
\betamod, the $\chi^2$ becomes stable after the inner $\sim 1\arcmin$
is excised; the fit results give $\beta = 0.65\pm0.04$ and $r_{\rm c}
= 1\arcmin.81^{+0.21}_{-0.20}$, with $\chi^2 = 78.52/60\ {\rm d.o.f}$.

Pratt \& Arnaud~(\cite{pa02}) discuss alternative parameterisations for
cases where a central excess is seen. These data were thus fitted
with their double isothermal \betamod\ (BB) and the generalised
double \betamod\ (KBB); the latter allowing a more centrally peaked gas
density profile The resulting best fit models are given
in Table~\ref{tab:sbfit}. Note that the profile obtained beyond
$R_{\rm cut}$ for the BB and KBB models is a classic \betamod.

The BB and KBB models can be compared using an F-test. The F-test
indicates that the KBB model is a better fit at the $42\%$ confidence
level, suggesting that the density distribution in the core is not
significantly more peaked than for a conventional \betamod. The BB
model is thus adopted for the remainder of this analysis.


\section{Temperature and abundance distribution}

\subsection{Preliminaries}

In the following, source and background spectra were extracted from
the weighted event lists. In the case of the pn, these spectra were
also corrected for out of time events, although in practice this
makes little difference to the results. The spectra are fitted
between 0.3 and 6.0 keV (cluster emission is not detected above
background beyond the upper limit), and the following response
matrices were used: m1\_thin1v9q20t5r6\_all\_15.rsp (MOS1),
m2\_thin1v9q20t5r6\_all\_15.rsp (MOS2) and epn\_ef20\_sY9\_thin.rsp
(pn).

\begin{figure}
\begin{centering}
\includegraphics[scale=0.5,angle=270,width=\columnwidth,keepaspectratio]{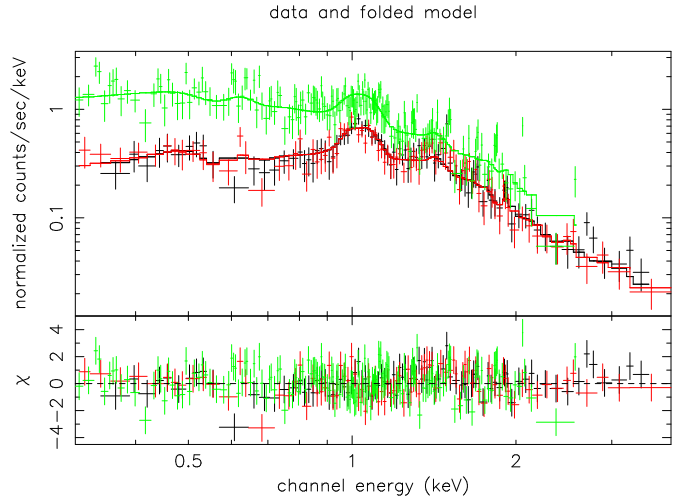}

\caption{{\footnotesize The global spectrum of A1983: MOS1 (black),
MOS2 (red) and pn (green), shown with the best fitting combined model
at $kT = 2.13\pm 0.08$ keV with $Z = 0.40 \pm 0.05
Z_{\odot}$.}}\label{fig:globspec}
\end{centering}

\end{figure}

\subsection{Global spectrum}
\label{sec:globspec}
As a consistency check between the three cameras, a global spectrum
was extracted from each event file using all events within the radius
of detection (8\arcmin.4, as established in Sect.~\ref{sec:gasden}).
Each spectrum was fitted with an absorbed MEKAL model with the
redshift fixed at $z=0.0442$, the mean redshift of the cluster.  After
checking that the fit results agreed when the $N_{\rm H}$ was left
free, the absorption was fixed at the galactic value of $N_{\rm H}=
2.04 \times 10^{20}$ cm$^{-2}$ (Dickey \& Lockman~\cite{dl90}); the
free parameters of each fit were thus the temperature and the
abundance (with respect to the solar abundances given by Anders \&
Grevesse~\cite{agr}).  The global spectral fits yield
$2.25^{+0.33}_{-0.12}$ keV, $2.06^{+0.13}_{-0.14}$ keV and
$2.06^{+0.17}_{-0.18}$ keV for MOS1, MOS2 and pn, respectively, values
which are in excellent agreement.  The combined MOS/pn fit yields $\kT
= 2.13\pm 0.08$ keV and $Z = 0.40 \pm 0.05 Z_{\odot}$.

The spectra are well fitted down to 0.3 keV. This is in contrast to a
previous \xmm\ observation of A1413 (Pratt \& Arnaud~\cite{pa02}), where,
below $\sim 0.6/1.0$ keV (MOS/pn) a marked excess was seen.  This
could indicate that the excess in A1413 may well be a true soft
excess, and not due to residual calibration problems.  The global
spectrum of A1983 is shown with the best fit model in
Fig.~\ref{fig:globspec}.

\subsection{Radial temperature profile}
\label{sec:temprof}
A radial temperature profile was produced by excluding point sources
and extracting spectra in circular annuli centred on the peak of the
X-ray emission. The widths of the rings were chosen so that a minimum
$5 \sigma$ detection in the [2-5] keV band was reached. This was
possible for all but the final annulus. A minimum width of 30\arcsec\
was also imposed, corresponding to the $90\%$ encircled energy radius
of the MOS PSF. The spectra were binned to $3\sigma$ above background
level to allow the use of Gaussian statistics.

\begin{figure}
\begin{centering}
\includegraphics[scale=1.0,angle=0,width=\columnwidth,keepaspectratio]{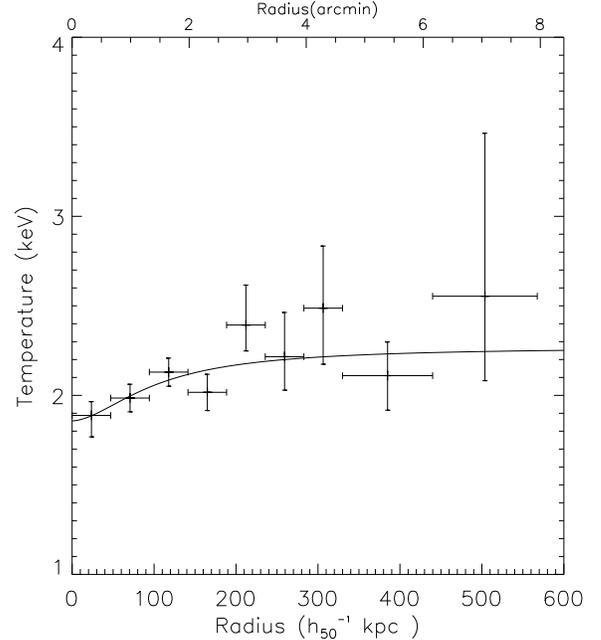}
\caption{{\footnotesize The total XMM-EPIC radial temperature profile
of A1983. The solid line is the best fit to the temperature profile
with a function of the form given in Eqn.~\ref{eqn:allen}. Errors are
$1 \sigma$ confidence level. }}\label{fig:annspec}
\end{centering}
\end{figure}

The spectra were fitted with the absorbed MEKAL model described in
the previous Section. In all cases the MOS normalisations were tied
together, but the pn normalisation was allowed to vary. In the final
two annuli, the abundance was frozen at the abundance found for the
third-last annulus.  Spectra from different cameras were fitted
individually at first: these results are identical within their
errors, and so a further simultaneous MOS+pn fit was undertaken. The
results are detailed in Table~\ref{tab:kt} and shown in
Fig.~\ref{fig:annspec}.

\begin{table}
\centering
\center
\caption{{\small Projected radial temperature profile results. Errors are
$1\sigma$, as shown in Fig.~\ref{fig:annspec}.}}
\begin{tabular}{ l l l}
\hline

\multicolumn{1}{l}{ Annulus } & \multicolumn{1}{l}{ kT } &
\multicolumn{1}{l}{$Z$} \\
\multicolumn{1}{l}{ ( \arcmin\ )} & \multicolumn{1}{l}{ (keV) } &
\multicolumn{1}{l}{ ($Z_{\odot}$} \\

\hline

0.00-0.66 & $1.88^{+0.07}_{-0.12}$ & $0.56^{+0.07}_{-0.06}$ \\
0.66-1.32 & $1.99^{+0.08}_{-0.08}$ & $0.34^{+0.05}_{-0.04}$ \\
1.32-1.98 & $2.13^{+0.08}_{-0.08}$ & $0.37^{+0.06}_{-0.05}$ \\
1.98-2.64 & $2.02^{+0.10}_{-0.10}$ & $0.29^{+0.06}_{-0.05}$ \\
2.64-3.30 & $2.39^{+0.22}_{-0.15}$ & $0.29^{+0.09}_{-0.07}$ \\
3.30-3.96 & $2.21^{+0.25}_{-0.19}$ & $0.29^{+0.10}_{-0.08}$ \\
3.96-4.62 & $2.49^{+0.35}_{-0.31}$ & $0.24^{+0.12}_{-0.09}$ \\
4.62-6.16 & $2.11^{+0.19}_{-0.19}$ & 0.3 frozen             \\
6.16-7.94 & $2.55^{+0.91}_{-0.47}$ & 0.3 frozen             \\

\hline
\end{tabular}
\label{tab:kt}
\end{table} %

\subsection{Is there a cooling flow?}
\label{sec:cf}

The slight and gradual decrease in temperature towards the centre
indicates that there might be a cooling flow (CF).  Indeed this has
already been suggested, on the basis of a deprojection analysis of the
Einstein data by White, Jones \& Forman (\cite{wjf97}), who find
evidence for a weak CF of $\sim 6 M_{\odot}$ yr$^{-1}$.  The cooling
time can be calculated using

\begin{equation}
t_{\rm cool} = 2.9 \times 10^{10}\ {\rm yr}\ \sqrt{\frac{\kT}{1~\keV}}
\left( \frac{n_{\rm H}}{10^{-3} {\rm cm}^{-3}} \right)^{-1},
\end{equation}

\noindent from Sarazin~\cite{cs86}. With the central density derived
from the BB model fit, $t_{\rm cool} \sim 8.5 \times 10^{9}$ yr. This
value suggests that cooling, if indeed present, should be rather weak.

Starting from the central annulus and working outward, the spectra
were fitted with an absorbed two temperature MEKAL model.  Except for
the central [0\arcmin - 0\arcmin.66] spectrum, the addition of a second
spectral component either does not provide a significant (as defined
by application of the F-test) reduction in the $\chi^2$, or, more
often, the second component parameters are completely unconstrained.

The  [0\arcmin - 0\arcmin.66] spectrum was then fitted with various
models as detailed in Table~\ref{tab:censpec}. The best-fitting model
is (very marginally) the absorbed MEKAL+MKCFLOW. The same spectrum
fitted with a single temperature plus cooling flow model with
$\dot{M}$ fixed to $6M_{\odot}$ yr$^{-1}$ (as obtained by White et
al.~\cite{wjf97}) yields $\chi^2_{\nu} = 1.93$ and shows significant
residuals below 1 keV.

\begin{table*}
\begin{minipage}{\textwidth}
\centering
\center
\caption{{\small Comparison of spectral fits to the  [0\arcmin -
0\arcmin.66] spectrum.  The $N_{\rm H}$ has been fixed to the galactic
values as detailed in Sect~\ref{sec:globspec}.  All errors are given
at the $90\%$ confidence level.}}
\begin{tabular}{ l l l l l l}
\hline

\multicolumn{1}{l}{ Model } & \multicolumn{1}{l}{ kT } &
\multicolumn{1}{l}{$Z$} & \multicolumn{1}{l}{$kT_{\rm low}$} &
\multicolumn{1}{l}{$\dot{M}$} & \multicolumn{1}{l}{ }\\
\multicolumn{1}{l}{ } & \multicolumn{1}{l}{ (keV) } &
\multicolumn{1}{l}{ ($Z_{\odot}$) } & \multicolumn{1}{l}{ (keV) } &
\multicolumn{1}{l}{ ($M_{\odot}$ yr$^{-1}$) } & \multicolumn{1}{l}{
$\chi^2$/d.of}\\

\hline

wabs(mekal) & $1.88^{+0.12}_{-0.14}$ & $0.56^{+0.12}_{-0.13}$ & - &
-  & 331.4/307  \\
wabs(mekal+mekal)\footnote{In this fit the abundance of heavy
elements is linked between the two components} &
$2.18^{+0.19}_{-0.16}$ & $0.82^{+0.18}_{-0.08}$ &
$0.82^{+0.25}_{-0.17}$  & - & 330.4/305 \\
wabs(mekal+mkcflow)\footnote{The maximum temperature of the CF is
tied to the temperature of the thermal component, and the abundances
are linked between the two components. The low temperature cutoff of
the CF is found to be $0.63^{+0.27}_{-0.20}$ keV.} &
$2.32^{+0.32}_{-0.10}$ & $0.85^{+0.18}_{-0.10}$ & - &
$3.20^{+0.80}_{-0.79}$ &  330.1/305 \\

\hline
\end{tabular}
\label{tab:censpec}
\end{minipage}
\end{table*} %



\subsection{Projection and PSF effects}

PSF and projection effects will be a particular problem if radial
bins smaller than the $90\%$ encircled energy radius of the PSF are
used, or if the surface brightness of the cluster is very peaked (i.e.,
the luminosity is concentrated towards the centre), or if there are
significant temperature gradients. None of which is true for A1983:
the radial temperature profile, extracted in bins with width
greater than the $90\%$ encircled energy radius of the PSF, is
relatively flat out to the limit of detection, and, while the surface
brightness profile does show a slight peak towards the centre, this
is not a large effect. In this context A1983 can be compared directly
with A1413, a cluster with a relatively flat temperature profile and
a peaked surface brightness distribution. The fully PSF-corrected and
deprojected temperature profiles of A1413 are consistent with the
projected profile within the $1\sigma$ errors. Since deprojection
and/or PSF correction are likely to have little effect on the
temperature profile, except to decrease the $S/N$, the projected
profile is used throughout this analysis.

\begin{figure}
\begin{centering}
\includegraphics[scale=1.0,angle=0,width=\columnwidth,keepaspectratio]{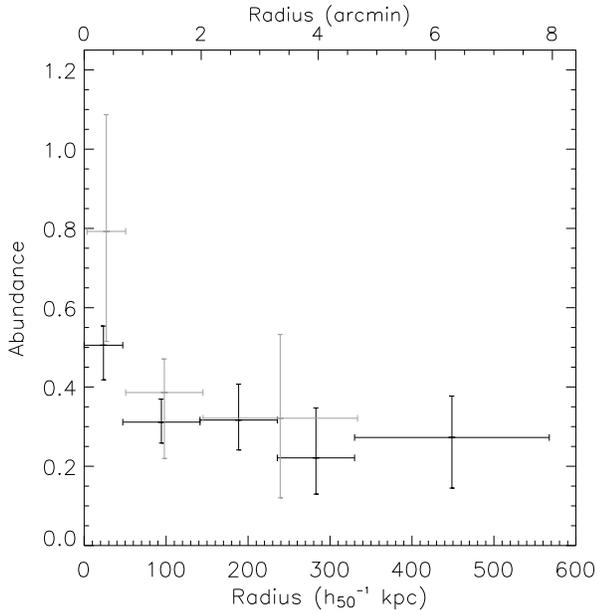}

\caption{{\footnotesize The radial abundance profiles for Fe (black)
and Si (grey).}}\label{fig:abprof}
\end{centering}
\end{figure}

\subsection{Abundance gradients}
\label{sec:abprof}

There is strong line emission from several elements at the average
temperature of A1983, however, significant radial constraints could
only be put on Fe and Si abundances.  (The Mg K line falls in the
energy band affected by the Al K and Si K flourescent energy lines in
the MOS background, and so was ignored in the present analysis.)

The radial abundance profile of Fe shown in Fig.~\ref{fig:abprof} was
obtained from an absorbed VMEKAL model fit with all $\alpha$-processed
elements and Fe treated as free parameters.  To augment the
statistics outside the centre, the spectra were reaccumulated 
in annuli twice the width
as those used for the temperature determination.  The Fe gradient is
relatively flat except for the central annulus, where there is a
pronounced jump from $0.3 Z_{Fe,\odot}$ to $0.5 Z_{Fe,\odot}$.

The spectra had to be reaccumulated in bins twice as large again for
constraints to be found on the Si profile, shown in the same Figure.
The Si abundance jumps from $0.4 Z_{Si,\odot}$ to $0.8 Z_{Si,\odot}$
in the central annulus, but outside this region the Si abundance is
essentially constant. Because of the large errors on the Si
abundance, it is difficult to obtain useful constraints on the radial
variation of the Si/Fe abundance ratio.


\subsection{Hardness ratio image}
\label{sec:2danalysis}

A hardness ratio (HR) image offers a relatively simple way of
analysing the two-dimensional temperature structure of the cluster,
and to this end, source and background images were produced in the
[0.3-0.9] keV and [2.5-6.5] keV bands, chosen to avoid strong line emission.
Exposure maps were generated
for each image in each band using the SAS task {\tt eexpmap}.

To avoid large numbers of negative pixels, especially in the external
regions, it is preferable to smooth images before undertaking any
mathematical operations. The SAS task {\tt asmooth} allows a
smoothing template to be defined, ensuring that each region of each
image can be smoothed in exactly the same fashion. Since the high
energy images have a lower sigal to noise ratio, the template is best
defined using these images. For the present analysis, a template was
defined by adaptively smoothing the MOS1 [2.5-6.5] keV image, and
this template was then applied to each source and background image.

\begin{figure}
\begin{centering}
\includegraphics[scale=1.,angle=0,width=\columnwidth,keepaspectratio]{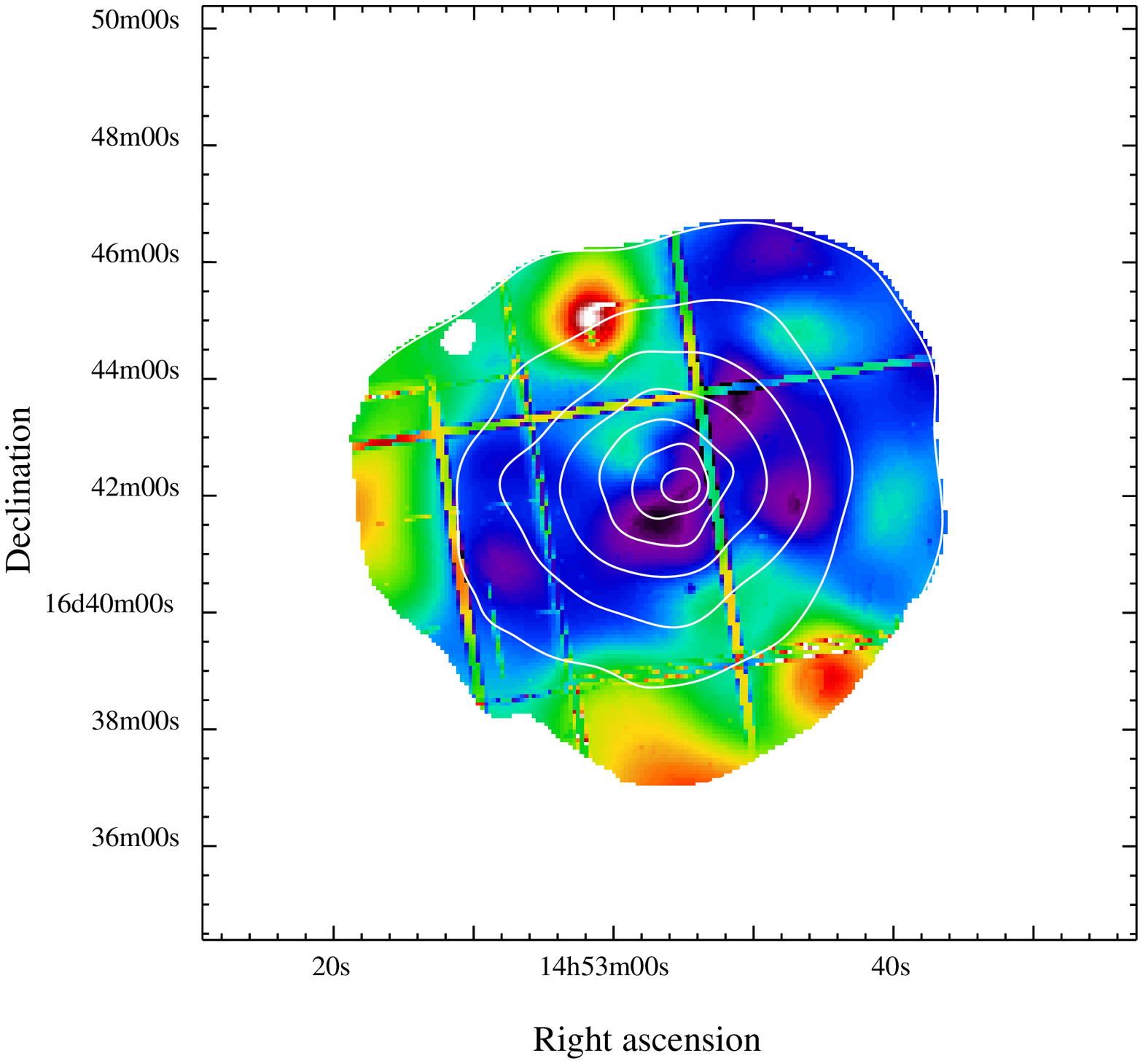}
\includegraphics[scale=1.,angle=270,width=\columnwidth,keepaspectratio]{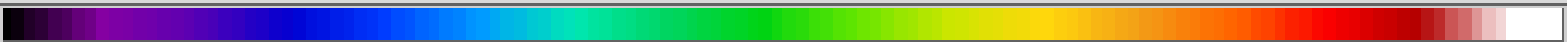}
\caption{{\footnotesize The hardness ratio map of the cluster
extending to a radius $\sim 7 \arcmin$ from the centre, from all EPIC
cameras. The dynamic scale has been chosen to emphasise structure, is
linear, and ranges from -0.75 (black) to -0.375 (white).
The lines are due to the gaps between CCD chips.}}\label{fig:hreqw}

\end{centering}
\end{figure}

The corresponding background was normalised based on the effective
exposure times and subtracted from each image. The background
subtracted images were then divided by their respective exposure
maps, and the hardness ratio image was calculated for each camera
using HR = (image[2.5-6.5 keV]-image[0.3-1.4 keV])/(image[2.5-6.5
keV]+image[0.3-1.4 keV]). The MOS and pn results are in excellent
agreement, and the combined MOS/pn HR image is shown in
Fig~\ref{fig:hreqw}.

In deriving these results, the difference between
the blank-sky background and the A1983 background at low energy 
has been neglected. This
would only affect the low-energy images as there is essentially no difference
above $\sim 2$ keV, and would preferentially affect the outer regions
where the background becomes a significant fraction of the detected flux.
In not taking this into account, the absolute HR values will be
incorrect. However, the overall temperature structure will still be
well described. Tests with the annular spectra (discussed above) show
that the difference between the backgrounds begins to play a role at
a radius of about 6\arcmin, making the uncorrected spectra seem
hotter than they actually are by about $10\%$. This would explain why
the external regions of the map seem slightly harder. A direct
translation of HR values into temperatures has not been attempted
because of the complexity properly of treating this problem.

The HR image shows quite a lot of structure, with point sources to
the NE and SW of the centre standing out clearly.  There appears to
be a cool elongation running approximately SE-NW, which has a radial
extension of $\sim 2 \arcmin$ (roughly 140 kpc). However, it should
be stressed that the colour table has been adjusted to emphasise any
structure, and in fact, the hardness ratio variations are not large
at all within the region covered by the map (as can be seen from the
colourbar). As a rough guide, the variations in hardness ratio present in
this image represent temperature variations of order $20\%$, a value which
is entirely consistent with the radial values seen in Sect.~\ref{sec:temprof}.

\begin{figure}
\begin{centering}
\includegraphics[scale=1.0,angle=0,width=\columnwidth,keepaspectratio]{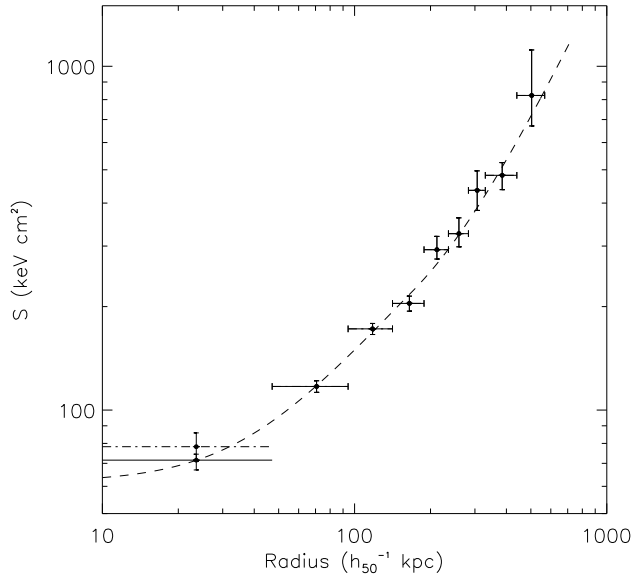}

\caption{{\footnotesize The gas entropy ($S=\kT/\ne^{2/3}$) profile
derived from the temperature profile and the double \betamod\ (BB)
for the gas
density.  The dotted line represents the entropy profile obtained
using analytic models for the gas density and temperature
distributions.  The dot-dash line is the entropy obtained in the
central bin if the spectrum is modelled with a two phase gas and the
hotter of the two components is used to calculate the
entropy.}}\label{fig:entropyprof}
\end{centering}
\end{figure}


\section{Entropy profile}

The gas entropy was determined from the BB analytical model fit to the
gas density profile and the observed temperature profile, with entropy
$S=\kT/\ne^{2/3}$.  The temperature profile was also modelled with a
function of the form:
\begin{equation}
T = T_0 +T_1 [(r/r_{\rm c})^{\eta} / (1 +
(r/r_{\rm c})^{\eta})]\label{eqn:allen}
\end{equation}

\noindent (c.f., Allen \etal~\cite{asf01}) with $T_0=1.86$ keV, $T_1 =
0.41$ keV, $r_c = 1\arcmin.46$ and $\eta=1.75$, and the entropy
calculated using this profile.  The resulting entropy profiles are
shown in Fig.~\ref{fig:entropyprof}.  The uncertainty in the entropy profile
is dominated by the uncertainty on the temperature distribution.
Typical errors, corresponding to the error in each bin of the
temperature profile are indicated in the figure.  As discussed above
in Sect.~\ref{sec:cf}, there is some evidence that the central
bin may have more than one temperature component.  The entropy in this
central bin calculated with the hotter of the two temperature
components is also shown in the Figure, it agrees with the single
temperature values within their errors.



\section{Mass analysis}
\label{sec:mass}

\begin{figure}
\begin{centering}
\includegraphics[scale=1.0,angle=0,width=\columnwidth,keepaspectratio]{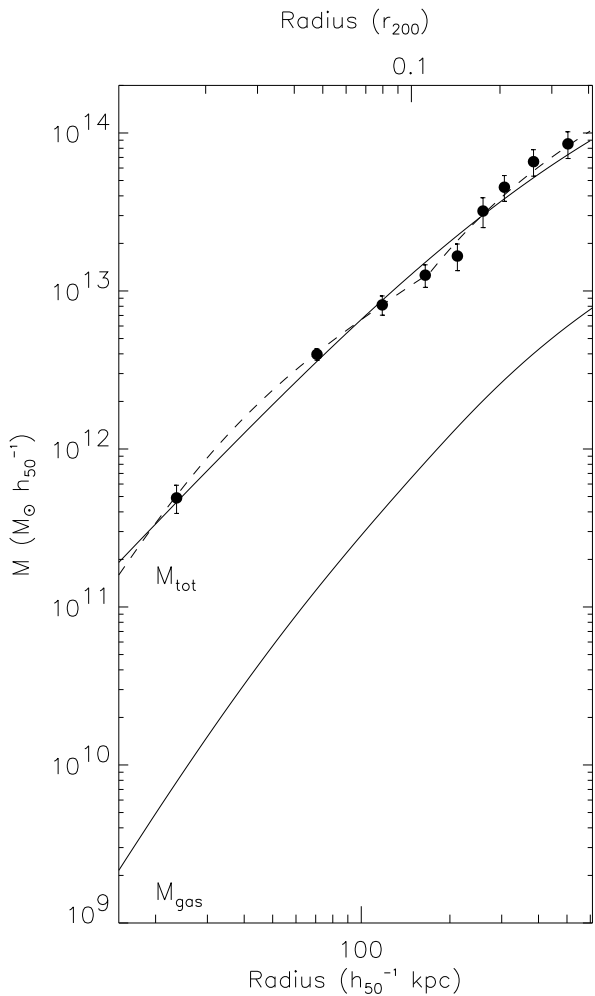}

\caption{{\footnotesize Integrated total gravitating mass and gas
mass distributions.  The total mass is shown with $1\sigma$ errors. 
The dashed line shows the mass profile calculated from the analytical 
temperature and gas density distributions. The solid line through the points
is the best 
fit NFW profile, with $c=3.75$ and $r_{200} = 1480$. In this and subsequent
Figures in Section~\ref{sec:mass}, this best-fitting NFW mass model is
used to compute the scaled radius given on the top axis.
}}\label{fig:massprof}
\end{centering}
\end{figure}
\subsection{Total mass profile}

\subsubsection{Calculation of the profile}

The total gravitating mass distribution shown in
Fig.~\ref{fig:massprof} was calculated under the usual assumptions of
hydrostatic equilibrium and spherical symmetry using

\begin{equation}
M(r) = - \frac{kT\ r}{{\rm G} \mu m_p}   \left[ \frac{d \ln{n_{\rm
g}}}{d \ln{r}} + \frac{d \ln{T}}{d \ln{r}} \right],
\label{eq:HE}
\end{equation}

\noindent where G and $m_p$ are the gravitational constant and proton
mass and $\mu = 0.609$.  The profile itself was calculated with an
adapted version of the Monte Carlo method of Neumann \&
B\"{o}hringer (\cite{nb95}), which takes as input the analytical
parameters for the gas density profile and the measured temperature
profile.  The original method calculates random temperature
distributions within the error bounds of the observed temperature
profile, using a `diffusive' process characterised by two parameters,
a window size and a step parameter.  A significant advantage of the
method is that it guarantees a certain smoothness of the temperature
profiles and thus naturally limits `unphysical' profiles with large
oscillations.  However, for this cluster, probably due to the
fluctuations observed in the temperature profile (see also below), it was
found difficult to choose objectively the window and step
parameters, on which depends the final dispersion on the derived mass
profiles and thus the error on the mass profile.  To avoid this
problem, a simpler, although less efficient, way to generate random
temperature profiles was used.  A random temperature at each radius of
the measured temperature profile was simply generated assuming a
Gaussian distribution with sigma equal to the $1\sigma$ error, and a
cublic spline interpolation was used to compute the derivative.  As in
Neumann \& B\"{o}hringer (\cite{nb95}) only `physical' temperature
profiles are kept, i.e. those yielding to monotically increasing total
mass profiles.  In total 1000 such profiles were calculated.

To include the error due to the uncertainty on the modelling of the
gas density profile, the error on the density gradient (i.e., ${\rm d}
\ln{n_{\rm g}} / {\rm d} \ln{r}$) was calculated.  For this, the
surface brightness profile was fitted at each radial point with ${\rm
d} \ln{n_{\rm g}} / {\rm d} \ln{r}$ as the free parameter, enabling
the errors on this quantity to be calculated from the $\chi^2$
variation, the other parameters (normalisation, $r_{c,{\rm in}}$,
$r_c$) having been optimised.  The final uncertainty values for the
mass points are derived from the quadratic addition of these errors
with those due to the temperature profile (output from the Monte Carlo
method).  The Monte Carlo mass profile, with appropriate
uncertainities, is shown in Fig.~\ref{fig:massprof}.

The mass profile can also be derived analytically using the
temperature profile model as given in Eqn.~\ref{eqn:allen} and the
best fit BB model.  This result, shown as the dashed line in
Fig.~\ref{fig:massprof}, is in excellent agreement with the results
from the Monte Carlo analysis.

\subsubsection{Modelling of the mass profile}

The total mass profile was fitted with the Navarro, Frenk \&
White~(\cite{nfw97}) profile, given by:
\begin{equation}
\rho(r) = \frac{\rho_{\rm c}(z) \delta_c}{(r/\rs) (1+ r/\rs)^2}
\end{equation}
\noindent where $\rho(r)$ is the mass density and $\rho_{\rm c} (z)$
is the critical density at the observed redshift, which, for a matter
dominated $\Omega = 1, \Lambda = 0$ Universe is:
\begin{equation}
\rho_{\rm c}(z) = {\rm \frac{3 H_0^2}{ 8 \pi G}} (1 + z)^3 ;
\end{equation}
\noindent and where $\delta_c$ can be expressed in terms of the
equivalent concentration  parameter, $c$:
\begin{equation}
\delta_{\rm c} = \frac{200}{3} \frac{c^3}{[\ln (1+c) - c/(1+c)]}.
\end{equation}
\noindent The radius corresponding to a density contrast of 200 is
$\rv= c\rs$.

This produced a fit with $\chi^2 = 9.14/7$ d.o.f, which is not a bad fit.
  The best fit NFW parameters are: $c = 3.75 \pm
0.74$, $\rs = 394 \pm 124$~kpc, giving $\rv = 1480$~kpc and $M_{200} =
2.15 \times 10^{14}$~M$_{\odot}$.  This best-fit NFW model is shown
overplotted on the mass profile of the cluster in
Fig.~\ref{fig:massprof}. Note that in this and other Figures in this Section,
the virial radius ($\rv$) shown is that calculated from the NFW fit.

\begin{figure}
\begin{centering}
\includegraphics[scale=1.0,angle=0,width=\columnwidth,keepaspectratio]{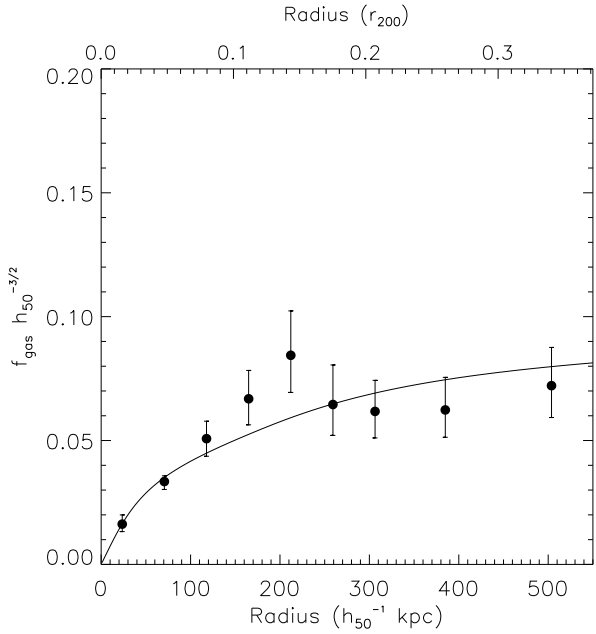}

\caption{{\footnotesize The integrated gas mass fraction as a function
of radius.  Errors are calculated from the quadratic combination of
$1\sigma$ errors on the total mass and the gas mass. The solid line is
the gas mass fraction derived from the gas mass and the best fitting
NFW mass profile.}}\label{fig:fgas}
\end{centering}
\end{figure}

The alternative Moore et al.~(\cite{mqgsl99}) profile was also tried,
which is given by:
\begin{equation}
\rho(r) = \frac{\rho_{\rm c}(z) \delta_c}{(r /r_s)^{3/2}
\left[1+\left(r/r_s \right)^{3/2}\right]},
\end{equation}
\noindent where
\begin{equation}
\delta_c = \frac{100 c^{3} }{\ln {(1 + c^{3/2})}}.
\end{equation}
With all parameters free, this profile gives a worse fit than the NFW
($\chi^2 = 12.22/7$ d.o.f), and moreover the best-fit scale
radius in this case is a rather unlikely 55 Mpc. The fit was repeated
with a upper limit to the scale radius fixed at twice the best-fit
NFW value (i.e., 788 kpc). Here the fit is worse again
($\chi^2 = 17.09/7$ d.o.f), but gives similar $\rv$ and $M_{200}$
values to the NFW fit.  The best-fit scale radius value is at the
maximum allowed by the fit.  In the light of these results, it seems
likely that these data do not have the required radial reach to put
useful constraints on this model.

\subsection{Gas mass profile and gas mass fraction}

Figure~\ref{fig:massprof} also shows the radial gas mass distribution
derived from integration of the best-fit BB model.
The variation in the gas mass fraction with radius is shown in
Fig.~\ref{fig:fgas}.  This figure shows that the gass mass fraction
rises rather abruptly in the first 200 kpc, beyond which it stabilises
at a value of $\sim 7\%$.

\begin{figure}
\begin{centering}
\includegraphics[scale=1.0,angle=0,width=\columnwidth,keepaspectratio]{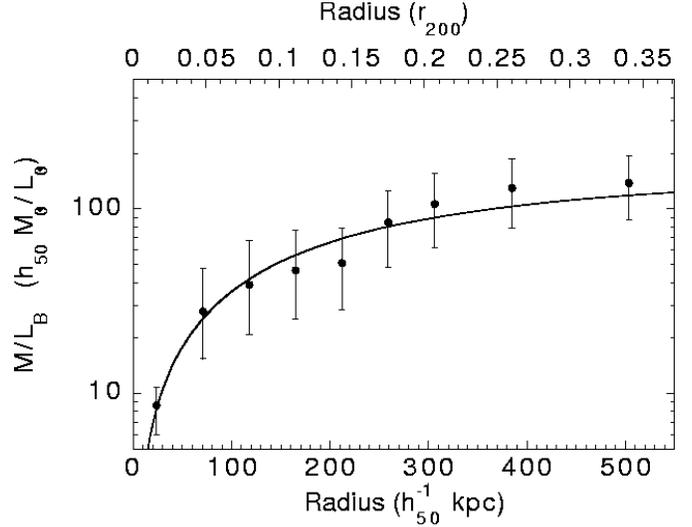}

\caption{{\footnotesize The integrated radial mass-to-light ratio.  Errors are
derived quadratically from the mass and luminosity errors in each
bin. The solid line is the mass-to-light ratio determined from the
best-fitting NFW mass model.}}\label{fig:mlratio}
\end{centering}
\end{figure}

\subsection{Mass to light ratio}
\label{sec:mlratio}

The extensive optical observations of A1983 can be used
to derive the radial mass-to-light ratio, which is
an indication of the extent to which the light traces the mass.

The contribution of the Central galaxy (cG) was the first to be
estimated.  The photometric data of Saglia \etal (\cite{saglia}) was
used: a blue magnitude of $B=16.0$ measured in an aperture of
$\theta_{\rm ap}=14.95\arcsec$, together with a surface brightness profile
following a de Vaucouleurs law with an effective radius of $\theta_{\rm
e}=14\arcsec$.  A total B magnitude of $B=15.3$ was derived, which
corresponds to a total blue luminosity of $L_{\rm B,cG} = 8.8 \times
10^{10}~\lsun$ for a galactic extinction of $A_{\rm B}=0.112$
(Schlegel \etal~\cite{schlegel}).  The luminosity within any radius of
interest was estimated using the aproximation of Mellier \& Mathez
(\cite{mellier}) for deprojecting the de~Vaucouleur surface brightness
profile.

Girardi \etal (\cite{getal02}) derive a total $B_{\rm j}$ luminosity
of A1983 from APS data.  After correcting for the completeness limit
they obtain $\lbj = 5.97/7.44\times10^{11}~{\rm L_{\odot}}$ within
$0.99$ Mpc for a Hubble constant of 100 km $s^{-1}$ Mpc$^{-1}$. 
The two luminosity
values are derived using different procedures for the fore/background
correction, so the average of the two values was used.  The luminosity
was then converted to the B band assuming $\lbj/{\rm
L_{B_{j},\odot}}=1.1~\lb/{\rm L_{B,\odot}}$ (Girardi
\etal~\cite{getal02}), and the luminosity of the cG subtracted.  The
resulting luminosity (excluding the cG) is then $\lb =
(2.4\pm0.2)\times10^{12}~\lsun$ within a projected radius of
$1.98~{\rm h_{50}^{-1}}$ Mpc.  A modified King profile with core
radius of $\rc = 20~{\rm h_{50}^{-1}}$ kpc and $\alpha$ = 0.61
(Girardi \etal~\cite{getal95}) can then be used to model the galaxy
distribution and thus interpolate the luminosity to the radius of
interest.  The $68\%$ errors on the profile parameters are large:
$\rc<100~{\rm h_{50}^{-1}}$ kpc and $0.5 < \alpha < 0.77$ (M. Girardi,
private communication), and are the main sources of uncertainty.  
This luminosity profile was then added to the cG
luminosity profile to derive the total cluster luminosity profile.


The final total mass-to-light ratio is shown in
Fig.~\ref{fig:mlratio}, where the errors are computed quadratically
from the mass and luminosity errors in each bin.

\subsection{ Iron mass over luminosity ratio}

The Iron Mass over Luminosity ratio (IMLR = $M_{\rm Fe}/\lb$) is a
fundamental quantity for the understanding of ICM enrichment (e.g.
Renzini~\cite{renzini}).  Using the Fe abundance measured at large
scale ($[Fe/H] \sim 0.3$ with typically $15\%$ error) the IMLR within
$0.3~\rv$ is IMLR $=(5.3\pm2.2)\times10^{-3}~\mlfsun$.  An overall IMLR
was also derived by extrapolating the gas mass to $\rv$ and assuming
that the abundance remains constant.  In this case, a great deal of
extrapolation is needed, and caution is advised in the interpretation
of the results.  On the other hand, the luminosity estimate within
that radius is more robust, being less sensitive to the assumed light
distribution.  The overall IMLR obtained using this method is IMLR
$=(8.9\pm1.8)\times10^{-3}~\mlfsun$.

\section{Discussion}

\subsection{Mass profile: comparison with theoretical expectations}

There is now converging evidence that the mass profile of relatively
hot clusters follows the cusped distribution predicted from numerical
simulations of dark matter collapse (David \etal~\cite{david01};
Arabadjis, Bautz \& Garmire~\cite{abg02}; Allen \etal~\cite{asf01},
Pratt \& Arnaud~\cite{pa02}).  Our results suggest that this is also
true for cool clusters.

The NFW profile is a relatively good fit to the A1983 data, but not as
good as for instance for A1413 (Pratt \& Arnaud~\cite{pa02}).
However, there are a several points of note.  For the calculation of
the mass profile, the radial temperature and gas density distributions
have been used.  The first and most obvious point is that the temperature
distribution is not smooth.  The radial temperature data used for this
mass analysis exhibit fluctuations on the order of $\sim 20\%$, and
this will inevitably have an effect on the quality of the NFW fit to a
mass model derived using these data.  Secondly, there is a point of
inflection in the mass profile around $170~{\rm h_{50}^{-1}~kpc}$,
corresponding to the boundary between the two $\beta$ models
used to fit the surface
brightness profile.  In the future, our intention is to develop more
robust models for the surface brightness modelling, but that is beyond
the scope of this paper.

The theory of cluster formation predicts that the concentration 
parameter should be higher for lower mass systems (e.g. NFW).  The
concentration parameter of A1983, $c=3.75\pm0.74$, is lower than would
be expected for a cluster of this mass, $c\sim 9$ (e.g Eke, Navarro \&
Frenk~\cite{enf}).  It is also significantly lower than the value
$c=5.4\pm0.2$ derived for A1413, a cluster about 5 times more massive
(Pratt \& Arnaud~\cite{pa02}).  However, studies of large numbers of
simulated haloes by Bullock~\etal (\cite{bull01}) and Thomas \etal
(\cite{pat01}) has shown that there is a wide dispersion in the value
of $c$, so the result is perhaps not surprising.  In addition, the
original NFW profile was derived from the mass distributions of
equilibrium haloes; in fact NFW specifically considered earlier epochs
for some of their simulated haloes, assuring equilibrium by excluding
the most recently accreted subunit.  As shown by Jing
(\cite{jing00}), in a study of the diversity of density profiles of a
large number of massive haloes, the quality of the NFW fit depends on
whether the halo is in equilibrium.  In particular, unrelaxed clusters
yield concentration parameters lower than the NFW values.
Observationally, a close look at the X-ray image of A1983
(Fig.~\ref{fig:amas})
indicates that, while the isophotes are regular at large scale, there
is an extension towards the NE in the central regions.  Moreover, as
the X-ray temperature analysis has shown (Sect.~\ref{sec:2danalysis}),
there are radial and non-radial temperature variations of $\sim 20\%$.
Lastly, previous optical observations
(Escalera \etal~\cite{escetal94}; Girardi \etal~\cite{getal97})
have found evidence for several substructures in the field of this
cluster, at least one of which appears to cover the central region.
These observational points seem to suggest that A1983 is not as
relaxed as initially it may appear, presumably due to an earlier
merging event which may still be detectable in the X-ray temperature
and galaxy distributions.  This shows the importance of knowing in
detail the dynamical state of a cluster for the interpretation of the
mass profile.

Large simulations, like the Hubble Volume simulation, indicate a
tight correlation between $M_{200}$ and the dark matter velocity
dispersion in the form of $\sigma = 1075~[(E(z)~M_{200}/(10^{15}~{\rm
h_{100}^{-1}~M_{\odot}})]^{1/3}$ km s$^{-1}$ (Evrard \& Gioia~\cite{eg}),
where $E(z)=H(z)/H_{0}$.  The best fit NFW mass model,
$M_{200}=2.15\times10^{14}~{\rm h_{50}^{-1}~M_{\odot}}$, gives $\sigma
\sim 520$ km~s$^{-1}$.  This is consistent with the robust, optically-derived
value of the galaxy velocity dispersion $\sigma = 551^{+71}_{-47}$
km~s$^{-1}$.  This excellent agreement indicates: i) that the total mass
estimate is probably not too much in error and that the cluster cannot
be too far from equilibrium, and ii) gives further support to the current
modelling of the dark matter collapse.

Finally, the normalisation of the $M_{200}$--$T$ relation for this
cluster, defined as $M_{10}(200)=E(z)~M_{200}~T_{10}^{-3/2}$, where 
$T_{10}$ is the temperature in units of 10 keV, is
$M_{10}(200) \sim 2.3\times10^{15}~{\rm h_{50}^{-1}~M_{\odot}}$. This is 
about $25\%$ lower than the normalisation of Evrard \etal (\cite{emn96}).
We thus confirm, at a lower mass, the offset between the observed and
theoretical normalisation of the $M$--$T$ relation, already observed
for hot clusters (Allen \etal~\cite{asf01}; Pratt \&
Arnaud~\cite{pa02}).


\subsection{Mass versus light as a function of scale}

At large radii, the light seems reasonably to trace the mass. Within the
relatively large errors, the mass-to-light ratio flattens 
beyond $\sim 0.2~\rv$, reaching a value of $\mlb =(135\pm45)~\mlsun$ 
at $0.3~\rv$ (Fig.~\ref{fig:mlratio}).  The value
extrapolated at $\rv$, using the best fit NFW mass profile, is
similar: $\mlb=150~\mlsun$, for a luminosity within that radius of
$\lb\sim1.4\times10^{12}~\lsun$ (estimated as described in
Sect.~\ref{sec:mlratio}).

\begin{figure*}
\begin{centering}

\includegraphics[scale=0.65,angle=0,keepaspectratio]{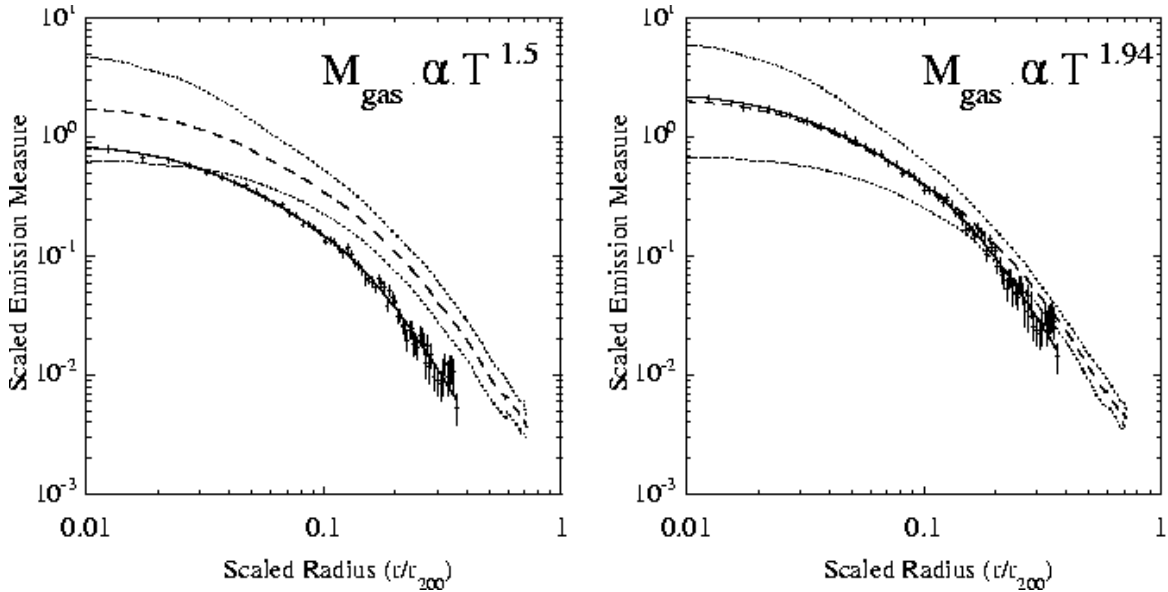}
\caption{{\footnotesize The scaled emission measure (EM) profile of
A1983 (data points) compared to the mean scaled EM profile of the
ensemble of hot clusters ($\kT > 3.5$ keV) observed by ROSAT and
considered by Neumann \& Arnaud (\cite{na01}).  The latter is
indicated by the dashed line; dotted lines show the dispersion.  On
the left, the EM profile of A1983 is scaled using the self-similar
\mgas$\propto \Tx^{1.5}$ scaling.  On the right, it is scaled using the
empirical \mgas$\propto \Tx^{1.94}$ scaling of Neumann \& Arnaud
(\cite{na01})}}\label{fig:emscaled}
\end{centering}
\end{figure*}

This mass-to-light ratio value for a cluster of such a low luminosity is
consistent with the general trend observed by Girardi \etal
(\cite{getal02}) from poor groups to rich systems.  Using a very large
sample, these authors found a significant increase of $M/L$ with
luminosity with a best fit power law of $M/{\rm M_{\odot}}
=10^{-1.596}~(\lb/{\rm L_{\odot}})^{1.338} $ for ${\rm
H_{0}}=100$~km~s$^{-1}$~Mpc$^{-1}$. Rescaling their results 
for ${\rm h_{50}} =1$,
and using the luminosity estimated at $\rv$, leads to a
mass-to-light ratio of $\mlb \sim (100\pm20)~\mlsun$, taking into
account the typical dispersion around the correlation.  This is within
the error bars of the observed value, although somewhat on the low
side.

On the other hand, David, Jones \& Forman (\cite{david}), who used
X-ray mass estimates for a much smaller sample of 7 groups and
clusters, did not find any evidence of mass-to light variation with
mass. All objects in their sample fall in the range $\mlv = 100-150 ~\mlsun$.
Similar values were derived by Cirimele, Nesci \& Trevese
(\cite{cirimele}) from a sample of 12 clusters in the temperature
range from 2 to 10 keV. Assuming $(L_{\rm V}/L_{\rm V,\odot}) =1.3~
(L_{\rm B}/L_{\rm B,\odot})$, relevant for early type galaxies, this
translates into a mass-to-light ratio in the B band of $\mlb= 130-200
~\mlsun$.  It must be noted that the mass-to-light ratio of A1983 is also
consistent with the David \etal (\cite{david}) finding.  Clearly
larger homogeneous X-ray samples, of size similar to those now available in
optical, are needed to see if there is a real contradiction between
$M/L$ trends derived respectively from X-ray and optical mass
determination methods.

The mass-to-light ratio decreases towards the central part of the
cluster, where the light is dominated by the central galaxy
(Fig.~\ref{fig:mlratio}).  The mass-to-light ratio in the central bin,
located at 23 kpc or about 1.4 times the effective radius of the 
central galaxy
(Sect.~\ref{sec:mlratio}), is $\mlb = 9\pm2$.  This value is consistent
with the typical mass-to-light ratio of the stellar population (e.g.
Gerhard \etal~\cite{gerhard} and references therein).  This suggests
that there is little if any dark matter component in the central part
of the galaxy.  At $r \sim 2.8 r_{e}$ the $\mlb$ ratio has increased to
a value of $\sim 30$.  Interestingly, similar values and trends have
been observed both around M87, the central very bright galaxy in the
Virgo cluster (Matsushita \etal~\cite{mat02}) and in isolated
elliptical galaxies (Gerhard \etal~\cite{gerhard}), two very different
environments.

\subsection{Scaling law tests}
\subsubsection{Testing the \mgas$-\Tx$ relation}

The surface brightness profile of A1983 can be used to test the
\mgas$-\Tx$ relation.  The surface brightness profile of the cluster
was converted into an emission measure ($EM$) profile using
\begin{equation}
EM(r) = \frac{4 \pi (1+z)^4 S(\theta)}{\Lambda(T,z)};\ \ r=d_{\rm A}(z),
\theta
\end{equation}
\noindent where $\Lambda(T,z)$ is the emissivity, taking into account
the interstellar absorption and the spectral response, and $r=d_A
(z)$ is the angular distance at redshift $z$.

This EM profile was then scaled according to the self-similar model,
using the standard scaling relations of cluster properties with
redshift and temperature\footnote{In all the Figures in this Section,
and for easy
comparison with previous works, the virial radius ($\rv$), 
is calculated using the normalisation given
in Evrard, Metzler \& Navarro~\cite{emn96}, viz., $\rv =
3.69~[\Tx/(10~\keV)]^{1/2}~(1+z)^{-3/2}~{\rm Mpc}$.  Note that when
comparing clusters of different temperatures, the exact value of this
normalisation does not matter.}.  In the first instance, the profile
was scaled using the standard relation \mgas$ \propto \Tx^{1.5}$.  In
the left-hand panel of Fig.~\ref{fig:emscaled}, this scaled EM profile
is compared to the mean scaled EM profile of the ensemble of hot ($\kT
>  3.5$ keV), nearby ($z < 0.06$) clusters observed by ROSAT and
considered by Neumann \& Arnaud (\cite{na01}).  It can clearly be seen
that the form of the profile is similar to that observed for hotter
clusters.  This is interesting, as is implies that the gas
distribution of this poor cluster is not more inflated than that
observed for hotter clusters.  In the BB model
(Sect.~\ref{sec:gasden}) the outer regions are modelled using a
standard \betamod: the best-fit $\beta$ value is
$\beta=0.76^{+0.13}_{-0.08}$, a value which is roughly consistent with (but
maybe a little higher than) the average value for hotter clusters,
$\beta=0.67$, found by Neumann \& Arnaud (\cite{na01}).

\begin{figure*}
\begin{centering}
\includegraphics[scale=0.5,angle=0,width=\columnwidth,keepaspectratio]{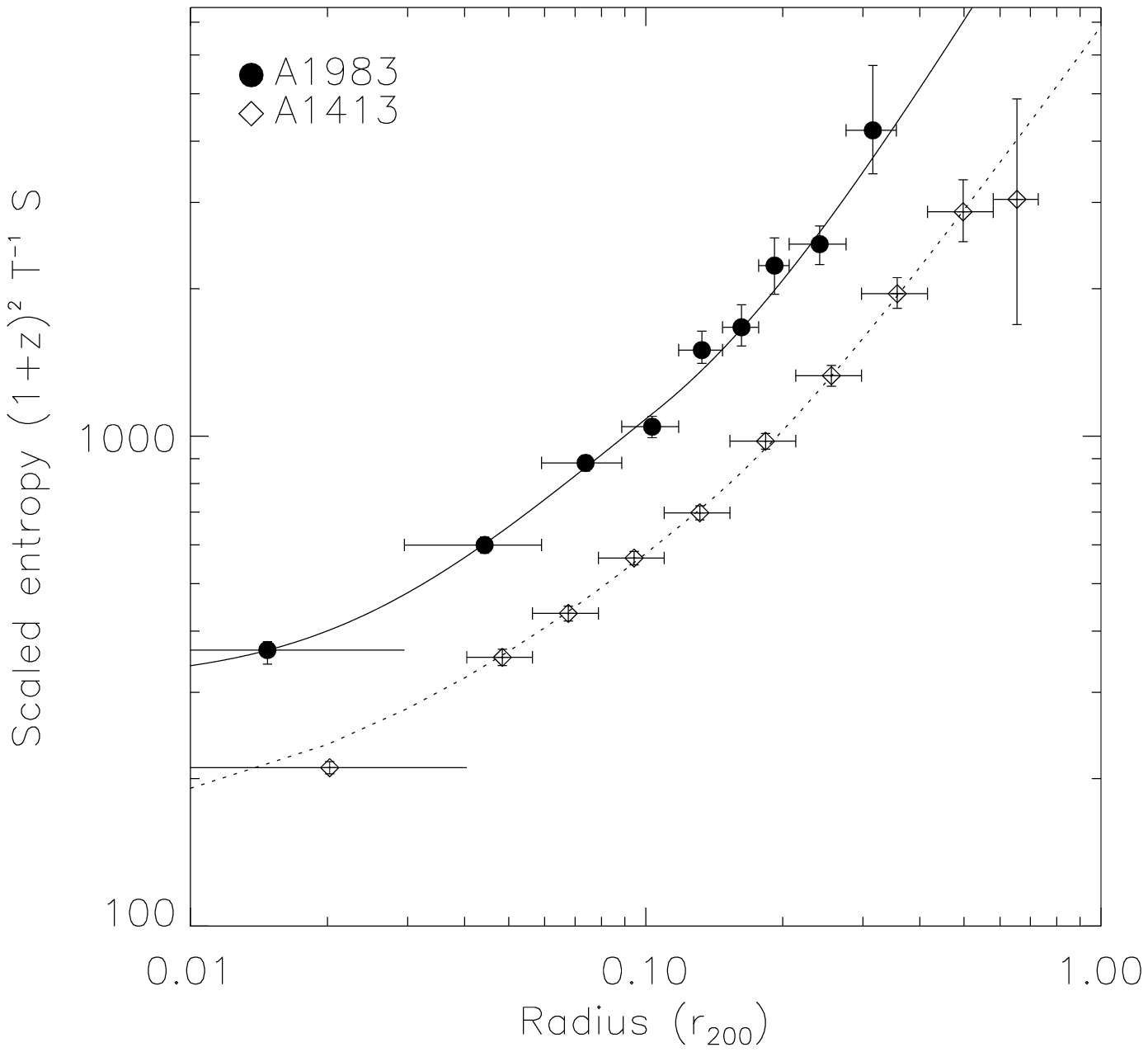}
\includegraphics[scale=0.5,angle=0,width=\columnwidth,keepaspectratio]{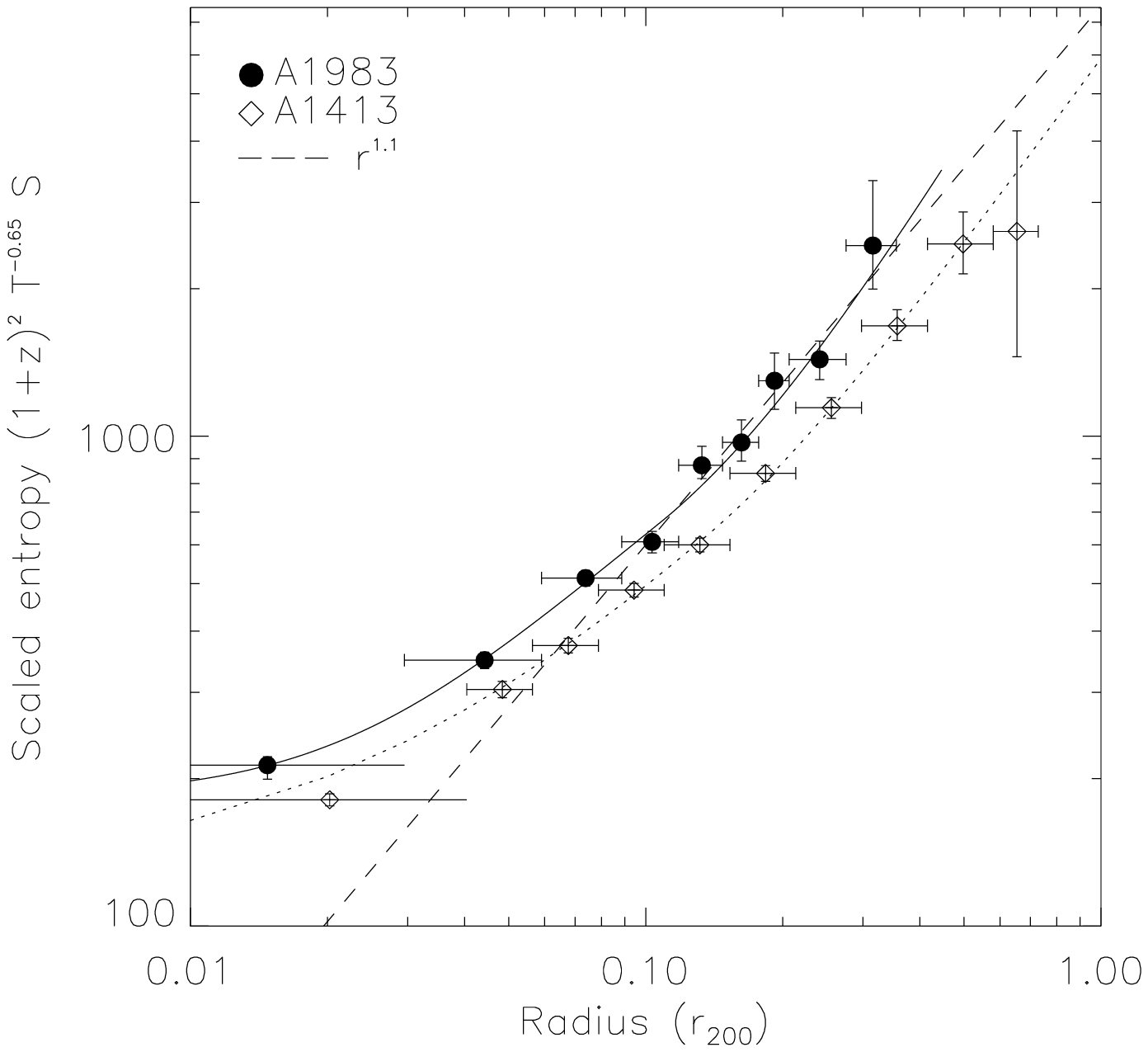}

\caption{{\footnotesize The scaled \xmm\ entropy profiles of A1983
and A1413. The points show the data. The solid line running
through the A1983 points is the entropy obtained when the temperature
profile is modelled using Eqn.~\ref{eqn:allen}; the dotted line
running through the A1413 points is the entropy obtained when the
temperature profile is modelled by a polytropic model with $\gamma =
1.07$ (see Pratt \& Arnaud \cite{pa02} for details). In the left hand panel,
the profiles are scaled with the self-similar $S \propto (1+z)^{-2} \Tx$ 
scaling. In the right panel, the profiles are scaled using $S \propto
(1+z)^{-2} \Tx^{0.65}$. The dashed line in the right-hand panel shows the $S
\propto r^{1.1}$ behaviour expected from shock heating, normalised to
the scaled entropy of a 10 keV cluster from Ponman
\etal~(\cite{psf03})}}\label{fig:enprof}
\end{centering}
\end{figure*}

However, there is an obvious problem with the normalisation (there is a
factor of two difference between the reference profile and that of A1983).
In the right-hand panel of Fig.~\ref{fig:emscaled}, the EM
profile has been scaled using the relation \mgas$ \propto \Tx^{1.94}$.
This relation, empirically-determined by Neumann \& Arnaud (\cite{na01}),
was found significantly to reduce the scatter in the scaled
profile of nearby hot clusters.  It is clear that using this \mgas$ -
\Tx$ relation, in other words, allowing a variable \fgas, has
considerably improved the normalisation, allowing A1983
to fall within the dispersion observed in the scaled EM profiles for
hotter clusters.

In summary, the gas distribution of this poor cluster is not more
inflated than that observed for hotter clusters, and the scaling of
the EM profile of A1983 is consistent with the steepening of the
\mgas$-\Tx$ relation observed for hotter clusters.

\subsubsection{The entropy profile and the $S - \Tx$ relation}

Here the \xmm\ entropy profile of A1983 is compared with
the \xmm\ entropy profile of the hot ($\kT=6.9~\keV$) cluster A1413
(for more details see Pratt \& Arnaud~\cite{pa02}).  In a self-similar
case, where \fgas\ is constant, $S \propto (1+z)^{-2}~\Tx$.  This
scaling is shown in the the left-hand panel of Fig.~\ref{fig:enprof}.
It can immediately be seen that the profiles are practically identical
in form.  This is itself is a remarkable result which will be
discussed in detail later on.

Ponman \etal (\cite{psf03}) have shown, with a sample of 66 virialised
systems, that the entropy measured at $0.1 \rv$ in fact scales as 
$S \propto \Tx^{\sim 0.65}$.
Implicit in this scaling is a dependence of the gas density $\ne$,
with system temperature.  Applying this scaling to the entropy
profiles of A1983 and A1413, shown in the right hand panel of
Fig.~\ref{fig:enprof}, shows that the normalisation between the two
profiles is considerably improved.  This Figure also shows the $S
\propto r^{1.1}$ behaviour, expected from analytical modelling of
shock heating in spherical collapse (Tozzi \& Norman~\cite{tn01}) and
normalised to the scaled entropy of a 10 keV cluster from Ponman
\etal~(\cite{psf03}).  It can be seen that overall, and down to
surprisingly small radius, both profiles agree remarkably well with
this prediction.  It is also clear that, although the profiles flatten
somewhat below about $0.1~\rv$, there is no evidence of an isentropic
core.


\subsubsection{Discussion}

The role of non-gravitational processes and the breaking of
self-similarity in clusters has long been a hot topic.  Approaches
have included analytic, semi-analytic and numerical modelling,
including pre-heating, internal heating by AGNs and SNs and/or cooling
(see Voit et al.~\cite{vbbb02} for a full review).

A dependence of \fgas\ on temperature has been predicted in some numerical
simulations (Dav\'e \etal~\cite{dave02}), and, as $\Lx \propto \ne^2$,
has been suggested as a principal cause of the similarity break in the
$\Lx$--$\Tx$ relation.  However, this variation has only been hinted
at in the observations (Mohr et al.~\cite{mme99} ; Arnaud \&
Evrard~\cite{ae99}).  Both of these papers determined global \fgas\
values using either \betamod\ or virial estimates for the total mass
of the system.  This approach, as stressed by the authors, is
obviously model-dependent.  The new \xmm\ data allow the derivation of
much more precise \fgas\ values, and allow the tracing of the \fgas\
with radius.

A1983 is obviously gas poor compared to the hotter cluster A1413.
This can be seen not only in the raw \fgas\ values (at $0.3 \rv$, for
A1983 $f_{\rm gas} = 0.08$ for A1983, while for A1413, $f_{\rm gas} =
0.16$) but also in the scaling behaviour of the emission measure
profiles.  Interestingly, Mushotzky et al.~(\cite{mfls03}) also find
remarkably low \fgas\ values in their analysis of \xmm\ data from two
galaxy groups (NGC2563 and NGC4325; gas mass fraction of $\sim 4-6 \%$
at $0.3 \rv$).  Thus there is strong evidence for some process,
proportionally more important in low mass systems, which either biases
intergalactic gas accretion, and/or removes gas from the hot X-ray
emitting phase of the ICM after collapse.  Note that the similarity of
the profiles over a wide temperature range indicates that the observed
variation of \fgas\ is not simply due to a displacement of gas from
the central regions to the outer regions.

Closely related to the question of a variation of \fgas, and
independent confirmation of that fact, is the entropy scaling
behaviour.  The relation $S \propto \Tx^{0.65}$, observed by Ponman
\etal (\cite{psf03}) acts to considerably improve the normalisation between
the entropy profiles.  This sort of modified $S$--$\Tx$ relation has been
seen in numerical simulations incorporating radiative cooling Dav\'e
\etal (\cite{dave02}).

Perhaps the most interesting aspect of the A1983 and A1413 entropy
profiles is their obvious similarity.  There is no isentropic core and
both profiles converge rapidly to the $S(r) \propto r^{1.1}$ value
favoured by models of the shock-dominated regime.  The scaled core
sizes are exactly the same.

As seen in e.g., Tozzi \& Norman (\cite{tn01}) and Babul \etal (
\cite{bblp02}), preheating models suggest that in the lowest mass
systems the accretion of the gas onto a potential well is isentropic,
with the shock heating regime becoming dominant at higher masses.  As
a result the entropy profiles of the lowest mass systems have large
isentropic cores, and these cores become progressively smaller with
increasing mass because of the increasing importance of shock heating.
This is clearly inconsistent with the present results (as also seen in
Ponman \etal~\cite{psf03} and Mushotzky \etal~\cite{mfls03}).

Given the evidence that preheating models cannot describe the
observed entropy profiles, one is forced to look towards models for
which the observational tests are much less evident. These models
invoke radiative cooling or heating by supernovae or AGN, or a
combination of the two, as a mechanism for modifying the entropy of
the ICM in lower mass systems.

Pure cooling models, while yielding results which are qualitatively
similar to the observations, seem to have problems with overcooling,
which furthermore is affected by the finite resolution of the
simulations (Muanwong et al.~\cite{muan02}; Dav\'e et
al.~\cite{dave02}).  It thus seems unlikely that pure cooling models
can work without the addition of some form of feedback of energy into
the baryonic component.  This feedback may be due to star formation,
or AGN heating, or any other plausible but undiscovered mechanism.
Moreover, there is a potential problem with the fate of the cooled
gas, which must not exceed the observed mass fraction of the stellar
component (e.g. Balogh \etal~\cite{bpbk01}; Dav\'e et al.~\cite{dave02};
Borgani \etal~\cite{borg02}).

An interesting analytical treatment is that of Voit \etal~(\cite{vbbb02}), 
where the problem is attacked using a
phenomenonological treatment of the entropy distribution of the gas.
The ICM entropy is either truncated such that all the gas with a
cooling time less than a Hubble time is removed (analagous to
radiative cooling), or shifted by raising the entropy level throughout
(analagous to preheating).  A generalised radiative loss model is also
used, which corresponds to a variable truncation of the entropy
distribution (analagous to some form of cooling with feedback).  In
common with other treatments, the preheated entropy distribution tends
to show a larger isentropic core, and also the pure radiative cooling
model seems to suffer from overcooling.  The authors suggest that
feedback is necessary to prevent this.

It thus seems clear that for cooling to be a viable alternative to the more
popular preheating scenario, some form of heating/feedback is
needed to prevent the overcooling.  The question now is how to
identify and distinguish between the competing processes, but from an
observational point of view it is unclear how this is to be done
without more stringent constraints from modelling and numerical
simulation.


\subsection{ICM enrichment}

It is instructive to compare the IMLR of A1983 with the ratio
observed in hotter clusters.  The results must however be treated
with some care, in view of the heterogenity of the various analyses.

From a compilation of the literature, Renzini (\cite{renzini}) found
a relatively small scatter in the IMLR down to a temperature of 2.2
keV, with a average value of $0.02~\mlfsun$.  The IMLR then drops suddenly
with temperature, to IMLR $< 0.01~\mlfsun$, for groups ($\kT <
1.8~\keV$).  With an IMLR of $8.9\pm1.8~10^{-3}$ (at $\rv$) for a
temperature of $2.1~\keV$, A1983 lies at the boundary of the two
regimes.  It must be noted however the IMLR values considered by
Renzini (\cite{renzini}) are not consistently estimated within the
cluster virial radii.  This is a potential source of bias, since the
IMLR is expected to increase with radius: whereas the Fe abundance
profiles are flat at large scale (De Grandi \& Molendi~\cite{dm01}),
there is evidence that the gas distribution is more inflated than the
galaxy distribution (e.g. Cirimele \etal \cite{cirimele}).  For
instance, above 3.5 keV, the data rely on the work of Arnaud \etal
(\cite{ar92}).  These authors estimated the IMLR within a fixed radius
of $3$~Mpc.  This corresponds to $r_{200}$ for a $\sim 6.6~\keV$
cluster.  The IMLR ratios of cooler (hotter) clusters are thus estimated
at a larger (smaller) fraction of the virial radius and thus are
probably over (under) estimated.  It is beyond the scope of this paper
to correct for this effect.  It should however be emphasized that once this
correction is made, one could well find a systematic increase of IMLR
with temperature, across the whole temperature range, and that the IMLR of
A1983 is clearly about two times less than for `hot' clusters.

An increase of IMLR from groups to rich clusters was observed by
Finoguenov, Arnaud \& David (\cite{fad}, Figure 8) for an IMLR
estimated consistently at a fixed fraction of the virial radius
($0.2$ and $0.4~\rv$).  Unfortunately the large error bars and considerable
scatter did not allow them to establish whether the IMLR systematically
increases with temperature or tends to a constant above some temperature
limit.  Again the IMLR ratio of A1983, $5.3\pm2.2~10^{-3}$ at
$0.3~\rv$, lies well within the observed trend and is again about 
two times smaller that the IMLR above $5~\keV$.

On the other hand the abundance of 0.3 is similar to the abundance
found in hot clusters (De Grandi \& Molendi~\cite{dm01}).  If
confirmed, such an increase of IMLR with system temperature (i.e mass)
while the abundance remains constant, will provide interesting information
on the evolutionary history of the gas.

A dependence on system mass of the Fe mass produced by unit of
stellar light is unlikely.  The abundance is set by the ratio of the
IMLR to the \mgas$/L$ ratio.  It would require an unlikely conspiracy 
to keep the abundance constant: a similar increase of the $IMLR$ and
the \mgas$/L$ ratio, quantities which are obviously not driven by the same
physics.  Furthermore, such a variable IMLR is unlikely on theoretical
grounds.  In view of the abscence of abundance evolution with $z$, it
is commonly assumed that the bulk of the Fe now seen in the
gaseous phase was produced in galaxies at early epochs (probably
before $z\sim 2$), and ejected through early galactic winds.  Neither
the heavy element production, determined by the chemical evolution of
the galaxies, nor the ejection efficiency are expected to vary with
system mass.

A more natural explanation of the {\it observed} variation of the IMLR
is that we do not see all the Fe mass ever produced and ejected by
the galaxies.  This issue is actually closely related to the question
of the variation of \fgas\ with system mass.  If the heavy elements
ejected by galaxies are homogeneously distributed in the intergalactic
medium, the X--ray determined IMLR$_{\rm X}$ will scale with 
the fraction of the gas with is now in the hot phase of the 
virialised part of the IGM, i.e with the gas mass fraction, \fgas, 
as determined from X-ray
observations.  On the other hand the abundance will rest constant.
Whatever mechanism explains an increase of \fgas\ with
T, the increase naturally implies a corresponding increase in IMLR$_{\rm X}$.

\section{Conclusions}

The main conclusions are summarised below.

\begin{itemize}

\item Results have been presented from the \xmm\ observation of the
poor cluster A1983 ($\kT = 2.1~\keV$), at $z=0.044$.

\item The gas density profile has been measured out to $\sim 8\arcmin.4$
and the temperature profile out to $\sim 7\arcmin$, or 
$\sim 500~{\rm h_{50}^{-1}~kpc}$, corresponding to
$\sim 0.35~\rv$.  The mass profile has been calculated out to the same
distance assuming HE and spherical symmetry.

\item The gas density profile is adequately modelled with a double
isothermal \betamod\ (BB), in which the gas density profile inside
and outside a radius $r_{\rm cut}$ is modelled by two different
\betamod s. The outer region \betamod\ parameter is $\beta = 0.74$.

\item The temperature profile exhibits a small drop towards the centre, 
but remains roughly constant out to the detection limit. Both the 
radial and
hardness ratio temperature information suggest fluctuations in 
temperature of $< 20\%$.
There is no evidence for multiphase gas
except in the very central annulus, where a two-temperature or a
cooling flow model gives a marginally better description of the data.

\item The mass profile is consistent with an NFW
profile with scale radius $\rs = 394$ kpc and a concentration
parameter $c = 3.75 \pm 0.74$.  A Moore \etal (\cite{mqgsl99}) profile
is unconstrained, due to the lack of data at large radii.  The low
value of $c$ may be indicative of the dispersion at lower masses observed 
in numerical simulations, or may suggest that the cluster is not 
completely relaxed. However, best fit NFW mass model gives a dark matter
velocity dispersion of $\sigma \sim 520$ km~s$^{-1}$, in excellent 
agreement with the optically derived galaxy velocity dispersion of 
$\sigma = 551^{+71}_{-47}$ km~s$^{-1}$. Finally, the normalisation of the 
$M_{200}-T$ relation for this cluster is $\sim 25\%$ too low with respect 
to the numerical simulations of Evrard et al.~(\cite{emn96}).

\item The gas mass fraction at $0.35 \rv$ is \fgas$= 8\%$.

\item When scaled using the self-similar relation \mgas$ \propto
\Tx^{1.5}$, the emission measure (EM) profile of A1983 lies a factor
of $\sim 2$ lower than the mean scaled profile of hot, nearby clusters
discussed in Neumann \& Arnaud (\cite{na01}), but has a similar shape.
If the empirically determined relation, \mgas$ \propto \Tx^{1.94}$, is
used instead, the scaled EM profile of A1983 matches well the
reference profile for hotter clusters.  This is direct evidence for a
variation of \fgas\ with temperature.

\item A direct comparison of the entropy profile of A1983 with the
\xmm\ entropy profile of A1413 ($\kT = 6.9~\keV$; Pratt \& Arnaud
2002) shows that the scaling between the profiles is much improved
when the empirically-determined relation $S \propto \Tx^{0.65}$ is
used, again supporting the suggestion of a variation of \fgas\ with
temperature.

\item The entropy profiles of A1983 and A1413 are remarkably similar
in form, with no evidence of an isentropic core, suggesting that
simple preheating models are not predicting the correct behaviour.
This is in agreement with recent \xmm\ and ASCA/ROSAT results,
favoring models which change the ICM entropy by internal heating,
cooling, or a combination of the two.

\item The mass-to-light ratio rises from $M/L_B = 9\pm 2 \mlsun$ 
at $23 h_{50}^{-1}$ kpc (typical of the 
mass-to-light ratio of the stellar population), to $M/L_B = 135 
\pm 45 \mlsun$ at $500 {\rm h}_{50}^{-1}$ kpc 
($0.35 r_{200}$; consistent with the trend of $M/L_B$ with system 
mass noted by previous optical observations).

\item The iron mass over luminosity within
$0.35~\rv$ is IMLR $=(5.3\pm2.2)\times10^{-3}~\mlfsun$. Extrapolating 
the gas mass to $\rv$ and assuming
that the abundance remains constant, the overall IMLR is 
$=(8.9\pm1.8)\times10^{-3}~\mlfsun$. The IMLR is about two
times less than the IMLR observed for clusters above 5 keV. This may 
also be connected to the variation of \fgas\ with system mass.

\end{itemize}

The results presented here have given some indication that the cluster
population remains self-similar in shape down to low temperature, that
the fundamental departure from the simplest self-similar model is an
increase of \fgas\ with temperature, and that there may indeed be some
dispersion in NFW fit parameters at the low mass end. Often in astronomy
however, the prototype is the exception.  In a forthcoming paper
the results presented here will be tested with a larger sample of
observations of poor clusters.

\begin{acknowledgements}
We thank M. Girardi for useful information on A1983 optical data and
D. Neumann for fruitful discussion on the determination of the mass
profile.  We also thank T. Ponman for very useful discussions and for
giving us a preprint of his paper. GWP is grateful to E. Belsole for
useful discussions. We are grateful to the anonymous referee for comments.
The present work is based on observations obtained with \xmm\, an ESA
science mission with instruments and contributions directly funded by
ESA Member States and the USA (NASA).  This research has made use of
the NASA's Astrophysics Data System Abstract Service; the SIMBAD
database operated at CDS, Strasbourg, France; the NASA/IPAC
Extragalactic database (NED) and the Digitized Sky Surveys produced at
the Space Telescope Science Institute.

\end{acknowledgements}

\end{document}